\documentclass[prd,superscriptaddress,unsortedaddress,twocolumn,showpacs,preprintnumbers,amsmath,amssymb]{revtex4}
\usepackage[dvips]{graphicx}
\usepackage{amsmath,amssymb,times}

\newcommand{\bequ}{\begin{equation}}
\newcommand{\eequ}{\end{equation}}
\newcommand{\bea}{\begin{eqnarray}}
\newcommand{\eea}{\end{eqnarray}}

\newcommand{\bfi}[1]{\mbox{\boldmath $#1$}}
\newcommand{\bfis}[1]{\mbox{\boldmath ${\scriptstyle #1}$}}

\newcommand{\vx}{{\bfi x}}
\newcommand{\vi}{{\bfi i}}
\newcommand{\vii}{{\bfis i}}

\newcommand{\vix}{{\bfis x}}

\DeclareSymbolFont{boldletters}{OML}{cmm} {b}{it}
\DeclareSymbolFontAlphabet{\mathbit}{boldletters}
\DeclareMathSymbol{\alpha}{\mathalpha}{letters}{"0B}
\DeclareMathSymbol{\beta}{\mathalpha}{letters}{"0C}
\DeclareMathSymbol{\gamma}{\mathalpha}{letters}{"0D}
\DeclareMathSymbol{\delta}{\mathalpha}{letters}{"0E}
\DeclareMathSymbol{\epsilon}{\mathalpha}{letters}{"0F}
\DeclareMathSymbol{\zeta}{\mathalpha}{letters}{"10}
\DeclareMathSymbol{\eta}{\mathalpha}{letters}{"11}
\DeclareMathSymbol{\theta}{\mathalpha}{letters}{"12}
\DeclareMathSymbol{\iota}{\mathalpha}{letters}{"13}
\DeclareMathSymbol{\kappa}{\mathalpha}{letters}{"14}
\DeclareMathSymbol{\lambda}{\mathalpha}{letters}{"15}
\DeclareMathSymbol{\mu}{\mathalpha}{letters}{"16}
\DeclareMathSymbol{\nu}{\mathalpha}{letters}{"17}
\DeclareMathSymbol{\xi}{\mathalpha}{letters}{"18}
\DeclareMathSymbol{\pi}{\mathalpha}{letters}{"19}
\DeclareMathSymbol{\rho}{\mathalpha}{letters}{"1A}
\DeclareMathSymbol{\sigma}{\mathalpha}{letters}{"1B}
\DeclareMathSymbol{\tau}{\mathalpha}{letters}{"1C}
\DeclareMathSymbol{\upsilon}{\mathalpha}{letters}{"1D}
\DeclareMathSymbol{\phi}{\mathalpha}{letters}{"1E}
\DeclareMathSymbol{\chi}{\mathalpha}{letters}{"1F}
\DeclareMathSymbol{\psi}{\mathalpha}{letters}{"20}
\DeclareMathSymbol{\omega}{\mathalpha}{letters}{"21}
\DeclareMathSymbol{\varepsilon}{\mathalpha}{letters}{"22}
\DeclareMathSymbol{\vartheta}{\mathalpha}{letters}{"23}
\DeclareMathSymbol{\varpi}{\mathalpha}{letters}{"24}
\DeclareMathSymbol{\varrho}{\mathalpha}{letters}{"25}
\DeclareMathSymbol{\varsigma}{\mathalpha}{letters}{"26}
\DeclareMathSymbol{\varphi}{\mathalpha}{letters}{"27}
\DeclareMathSymbol{\Gamma}{\mathalpha}{letters}{"00}
\DeclareMathSymbol{\Delta}{\mathalpha}{letters}{"01}
\DeclareMathSymbol{\Theta}{\mathalpha}{letters}{"02}
\DeclareMathSymbol{\Lambda}{\mathalpha}{letters}{"03}
\DeclareMathSymbol{\Xi}{\mathalpha}{letters}{"04}
\DeclareMathSymbol{\Pi}{\mathalpha}{letters}{"05}
\DeclareMathSymbol{\Sigma}{\mathalpha}{letters}{"06}
\DeclareMathSymbol{\Upsilon}{\mathalpha}{letters}{"07}
\DeclareMathSymbol{\Phi}{\mathalpha}{letters}{"08}
\DeclareMathSymbol{\Psi}{\mathalpha}{letters}{"09}
\DeclareMathSymbol{\Omega}{\mathalpha}{letters}{"0A}

\begin{document}
\preprint{SAGA-HE-285}
\title{Interplay between sign problem and $Z_3$ symmetry 
in three-dimensional Potts model}

\author{Takehiro Hirakida}
\email[]{hirakida@email.phys.kyushu-u.ac.jp}
\affiliation{Department of Physics, Graduate School of Sciences, Kyushu University,
             Fukuoka 819-0395, Japan}

\author{Hiroaki Kouno}
\email[]{kounoh@cc.saga-u.ac.jp}
\affiliation{Department of Physics, Saga University,
             Saga 840-8502, Japan}

\author{Junichi Takahashi}
\email[]{takahashi@phys.kyushu-u.ac.jp}
\affiliation{Department of Physics, Graduate School of Sciences, Kyushu University,
             Fukuoka 819-0395, Japan}

\author{Masanobu Yahiro}
\email[]{yahiro@phys.kyushu-u.ac.jp}
\affiliation{Department of Physics, Graduate School of Sciences, Kyushu University,
             Fukuoka 819-0395, Japan}

\date{\today}

\begin{abstract} 
We construct four kinds of $Z_3$-symmetric three-dimensional (3-d) 
Potts models, each with different number of states at each site 
on a 3-d lattice, by extending the 3-d three-state Potts model. 
Comparing the ordinary Potts model with the four $Z_3$-symmetric Potts models, 
we investigate how $Z_3$ symmetry affects the sign problem and see 
how the deconfinement transition line 
changes in the $\mu$-$\kappa$ plane as the number of states increases, 
where $\mu$ ($\kappa$) plays a role of chemical potential (temperature) 
in the models. We find that 
the sign problem is almost cured by imposing $Z_3$ symmetry. 
This mechanism may happen in $Z_3$-symmetric QCD-like theory. 
We also show that the deconfinement transition line 
has stronger $\mu$-dependence 
with respect to increasing the number of states. 
\end{abstract}

\maketitle

\section{Introduction}
\label{sec:intro}

Thermodynamic properties of quantum chromodynamics (QCD) 
are 
often described as a phase diagram in the $\mu$-$T$ plane, 
where $T$ and $\mu$ mean temperature and quark-number chemical potential, 
respectively. However, the QCD phase diagram is known only in the vicinity of 
the $T$ axis, because lattice QCD (LQCD) simulations 
as the first principle calculation 
have a serious sign problem at finite real $\mu$. 
Exploration of the QCD phase diagram then becomes 
one of the most challenging subjects 
in particle and nuclear physics, and 
the results have been playing an important role on cosmology and astrophysics. 
The quark determinant, which appears after the quark field is integrated out in the grand-canonical QCD partition function in the Euclidean path integral formalism, becomes complex for finite real $\mu$.  
Several approaches were proposed so far to resolve this sign problem; 
for example, the reweighting method~\cite{Fodor}, the analytic continuation from imaginary $\mu$ to real $\mu$~\cite{FP,D'Elia,D'Elia3,FP2010,Nagata,Takahashi} and the Taylor expansion method~\cite{Allton,Ejiri_density}. 
Recently, remarkable progress has been made in the complex Langevin simulation~\cite{Aarts_CLE_1,Aarts_CLE_2,Sexty,Greensite,Aarts_CLE_3} and the Picard-Lefschetz thimble theory~\cite{Aurora_thimbles,Fujii_thimbles,Tanizaki}, but our understanding are still far from perfection. 

$Z_3$ symmetry is exact in pure gauge theory but not in QCD with finite quark masses. 
However, the symmetry is considered to work well as an approximate symmetry and be related to the deconfinement transition. 
It was pointed out that $Z_3$ symmetry or group may play an important role in the sign problem.  
In Refs.~\cite{Condella} and \cite{Z3C}, it was shown that, using the properties of $Z_3$ group elements, 
an effective center field theory with sign problem can be transformed into a flux model with no sign problem. 
It was also conjectured that the center dressed quarks undergo a new phase due to the Fermi Einstein condensation in cold but dense matter of the hadronic phase and the sign problem is cured in such a new phase~\cite{Langfeld_center}.  
Recently, in Ref.~\cite{Z3A}, it was suggested that, even in the case of the full QCD having $Z_3$ symmetry approximately,  the sign problem may be cured to some extent by using the $Z_3$-averaged subset method, at least in the strong coupling limit. 
However, there are still many difficulties, 
when the methods are applied to realistic full LQCD simulations.  

As mentioned above, in pure SU(3) gauge theory, $Z_3$ symmetry is exact and 
characterizes the confinement-deconfinement transition. 
The Polyakov loop $L({\vx})$~\cite{Polyakov},  not invariant under the $Z_3$ transformation, is an order parameter of the transition. 
The expectation value $\langle L ({\vx})\rangle$ 
is also related with the free energy of a static quark. 
If $\langle L ({\vx}) \rangle=0$, the free energy becomes infinite 
and thereby quarks are confined. 
In the full QCD with dynamical quarks, however, $Z_3$ symmetry is not exact anymore and the relation between $Z_3$ symmetry and the confinement-deconfinement transition is not clear.  

In a series of papers~\cite{Kouno_TBC,Sakai_TBC,Kouno_adjoint,Kouno_TBC_2,Kouno_DFTBC}, $Z_3$-symmetric QCD-like theory was proposed and studied 
extensively. In this paper, we  simply refer to the theory as $Z_3$-QCD. 
In $Z_3$-QCD, the symmetric three flavor quarks are considered and the flavor dependent imaginary chemical potential is introduced to retain 
$Z_3$ symmetry. 
In the papers, the phase structure in the $Z_3$-QCD was studied by using 
the Polyakov-loop extended Nambu--Jona-Lasinio  (PNJL) model~\cite{Meisinger,Dumitru,Fukushima,Ratti,Megias} as an effective model of QCD. 
Recently, lattice simulations of $Z_3$-QCD was done 
at $\mu =0$~\cite{Iritani}.  
The result was shown to be consistent with the PNJL-model prediction. 
It should be also remarked that the $Z_3$-QCD tends to three flavor QCD 
in the limit of $T\to 0$ as is explained in next section.  

There is a possibility that the sign problem becomes milder in $Z_3$-QCD than 
in QCD.
Kouno {\it et al.} pointed out this possibility 
by making a qualitative comparison between the heavy-dense model~
\cite{Greensite:2014cxa,Aarts:2014kja} 
and its $Z_3$-symmetric extension newly constructed, 
and proposed a Taylor-expansion method 
of deriving QCD results from $Z_3$-QCD ones~\cite{Kouno_DFTBC}. 
The interplay between the sign problem and $Z_3$ symmetry is thus quite 
interesting.  

The heavy-dense model  is well known as an effective model of QCD for large mass $M$ and large $\mu$  keeping $\mu -M$ is finite. 
The three-dimensional (3-d) 3-state Potts model~\cite{Condella,Z3C,DeGrand,Karsch,Alford} is a simplified version of the heavy-dense model. 
In this sense, the Potts model should be considered at large $M$ and $\mu$.      
In the Potts model, three elements $(1, \exp[\pm i 2\pi/3])$ of the group $Z_3$ are taken as three states at each site on a 3-d lattice. 
Considering the $Z_3$ elements  as a substitute for the Polyakov-loop operator $L({\vx})$ in QCD, one can discuss $Z_3$ symmetry and confinement  through the 3-d Potts model. 
In QCD, the operator  $L({\vx})$ depends on the 3-d coordinate ${\vx}$ and $T$, 
but in the Potts model the $Z_3$ elements can depend on ${\vx}$ but lose 
information on $T$   as a result of the substitution. 
One can regard ${\vx}$-dependent $Z_3$ elements as `` 
Polyakov-loop operator" $\Phi_{\vix}$ in the  3-d Potts model. 
The $\Phi_{\vix}$ should be averaged over ${\vx}$ 
 to define random and ordered states. 
The average $\langle \Phi \rangle$ 
depends on the coupling strength $\kappa$ of 
the interaction between $\Phi_{\vix}$ at site ${\vx}$ and those 
at the  nearest neighbor sites.   
As an interesting result, $\kappa$ dependence of $\langle \Phi \rangle$ 
in the Potts model 
is shown to be similar to $T$ dependence of the Polyakov loop $\langle L({\vx}) \rangle$ in QCD~\cite{DeGrand}. 
This indicates that $\kappa$ plays a role of temperature and one can 
consider the phase diagram in the $\mu$-$\kappa$ plane. 

In this paper, we construct four kinds of $Z_3$-symmetric 3-d Potts 
models, each with different number of states, 
in order (1) to clarify the interplay between $Z_3$ symmetry and the sign problem, and 
(2) to see how the deconfinement transition line 
changes in the $\mu$-$\kappa$ plane 
with respect to increasing the number of states of $\Phi_{\vix}$. 
Basic properties of the four $Z_3$-symmetric 3-d Potts 
models, named (A)-(D), are tabulated in 
Table \ref{tb:models}, together with those of 
the original 3-d 3-state Potts model. 
We clarify subject (1) by comparing the original Potts model with model (D) 
and subject (2) by changing the number of states from 3 of model (A) to 
13 of model (D).

\begin{table}[b]
\begin{center}
\begin{tabular}{c|c|c|c}
\hline \hline
model & states & $Z_3$ symmetry & Free from sign problem? \\ \hline
Potts & 3 & non-symmetric & No \\
(A) & 3 & symmetric & Yes \\
(B) & 4 & symmetric & Yes \\
(C) & 7 & symmetric & Yes \\
(D) & 13 & symmetric & No \\ \hline \hline
\end{tabular}
\caption{Potts and $Z_3$-symmetric Potts models considered in this paper.
        $Z_3$-symmetric Potts models (A)-(D) are classified 
	with the number of states at each site. 
	The definition and properties of models are shown 
in Sec. \ref{Potts} for the Potts model 
and in Sec. \ref{Z3Potts} for the 
	$Z_3$-symmetric Potts models. 
      }
\label{tb:models}
\end{center}
\end{table}

The Potts model is missing the chiral symmetry, because it considers 
the case of large quark mass. Hence we cannot discuss 
the chiral phase transition directly. 
However, there is an onset transition of quark number density, 
in addition to the deconfinement transition, in the Potts model. 
The order parameter of the transition is, of course, the quark number density itself. As for the critical endpoint of chiral transition at finite $T$ and 
$\mu$ in QCD, 
it was pointed out ~\cite{Fujii} that the order parameter is not the 
scalar density only, but a linear combination of the scalar, quark number and energy densities. 
This mixing makes the chiral transition the first order 
at middle and large $\mu/T$. 
In this sense, we can regard the quark number density as the quantity 
related to the chiral transition at middle and large $\mu$. 
Similarly, the onset transition also affects the deconfinement transition. 
We will show that, in the $Z_3$-symmetric 3-d Potts model, the sign problem appears when the deconfinement transition is entangled strongly 
with the onset transition. 

This paper is organized as follows. 
We recapitulate $Z_3$-QCD in Sec.~\ref{Z3QCD}. 
In Sec.~\ref{real}, we overview  the sign problem and 
show a way of making the QCD partition function real.  
In Sec.~\ref{Potts}, the sign problem in the 3-d 3-state Potts model 
is examined. 
In Sec.~\ref{Z3Potts}, we construct $Z_3$-symmetric 3-d Potts models. 
Numerical simulations are done for the models in Sec.~\ref{Numerical}.  
Section~\ref{summary} is devoted to a summary. 

\section{QCD and $Z_3$-QCD}
\label{Z3QCD} 

As for finite $T$, 
the action $S$ of three-flavor QCD with symmetric quark masses 
($m$) and symmetric chemical potentials ($\mu$) is defined by 
\begin{eqnarray}
&&S=\int_0^\beta dx_4\int_{-\infty}^\infty d^3x {\cal L} 
\label{action}
\end{eqnarray}
with $\beta =1/T$ and the Lagrangian density  
\begin{eqnarray}
{\cal L}&=&{\cal L}_{\rm Q}+{\cal L}_{\rm G} 
\label{lagrangian} 
\end{eqnarray}
composed of the quark and gluon parts 
\begin{eqnarray}
{\cal L}_{\rm Q}&=&\bar{q}{\cal M}(\mu, A_\mu)q,
\label{lagrangian_Q} 
\\
{\cal L}_{\rm G}&=&{1\over{4g^2}}{F_{\mu\nu}^a}^2={1\over{2g^2}}{\rm Tr}\left[{F_{\mu\nu}}^2\right] , 
\label{lagrangian_G} 
\end{eqnarray}
where 
\begin{eqnarray}
{\cal M} (\mu,A_\mu) &=&D_\mu (A_\mu) \gamma_\mu +m-\mu\gamma_4, 
\label{eq:Delta}\\
F_{\mu\nu}&=&\partial_\mu A_\nu-\partial_\nu A_\mu-ig[A_\mu,A_\nu], 
\label{Fmunu}\\
A_\mu &=&\sum_{a=1}^8A_\mu^aT^a. 
\label{Anu}
\end{eqnarray}
Here $q$, $A_\mu$ and $T^a$ are the quark field, the gluon field, 
and the generator of SU(3) group, 
respectively, and the trace ${\rm Tr}$ is taken for the color indices. 
As for $T^a$, we use the standard notation $T^a={\lambda^a\over{2}}$ defined by the Gell-Mann matrices $\lambda^a$.  
The action $S$ then can be decomposed into the 
quark and gluon components as 
$S=S_{\rm Q}+S_{\rm G}$ with 
\begin{eqnarray}
S_{\rm Q}=\int_0^\beta dx_4\int_{-\infty}^\infty d^3x {\cal L}_{\rm Q},~
S_{\rm G}=\int_0^\beta dx_4\int_{-\infty}^\infty d^3x {\cal L}_{\rm G}. 
\label{action_G}
\end{eqnarray}

In the present formalism, we have used the following Euclidean notations: 
\begin{eqnarray}
&&x_4\equiv  ix_0=ix^0=it,~A_4\equiv iA_0=iA^0,~~~~~
\nonumber\\
&&D_\mu(A_\mu)={\partial\over{\partial x_\mu}}-iA_\mu,~~~~
\nonumber\\
&&\gamma_4=\gamma^0_{\rm M}, \gamma_i\equiv i\gamma^i_{\rm M}~(i=1,2,3),  
\label{notation}
\end{eqnarray}
where the $\gamma^\mu_{\rm M}~(\mu =0,1,2,3)$ are the standard gamma matrices in Minkowski space~\cite{Bjorken}.  
The temporal anti-periodic boundary condition on the quark field is given by 
\begin{eqnarray}
q(x_4=\beta,{\vx})=-q(x_4=0,{\vx}), 
\label{boundary_q_u}
\end{eqnarray}
while the gluon field satisfies the temporal periodic boundary condition.  
The Lagrangian density $\cal{L}$ is invariant under the $Z_3$ transformation, 
\begin{eqnarray}
q&\rightarrow &q^\prime =Uq, \nonumber\\
A_\mu &\rightarrow & A_\mu^\prime =UA_\mu U^{-1} +i(\partial_\mu U)U^{-1}, 
\label{Z3_trans}
\end{eqnarray}
where 
\begin{eqnarray}
U(x_4=\beta, {\vx})=\exp{(i\alpha_aT_a)}
\label{U_Z3}
\end{eqnarray}
is an element of SU(3) group characterized by real functions $\alpha_a$ satisfying the boundary condition 
\begin{eqnarray}
U(x_4=\beta, {\vx})=\exp{(-i2\pi k/3)}U(x_4=0,{\vx})  
\label{bc_U}
\end{eqnarray}
for any integer $k$. 
However, the quark field boundary condition (\ref{boundary_q_u}) is changed by the $Z_3$ transformation as 
\begin{eqnarray}
q(x_4=\beta,{\vx})=-e^{i2\pi k/3}q(x_4=0,{\vx}). 
\label{boundary_q_u_change}
\end{eqnarray}
In full QCD with dynamical quarks, $Z_3$ symmetry is thus broken through the quark boundary condition. 

$Z_3$ symmetry can be recovered by introducing the flavor-dependent twist boundary condition (FTBC) as follows. 
\begin{eqnarray}
q(x_4=\beta,{\vx})=-e^{-i\theta_f}q(x_4=0,{\vx})
\label{FTBC}
\end{eqnarray}
with 
\begin{eqnarray}
\theta_f={2\pi\over{3}}f~~~~~(f=-1,0,1). 
\label{theta_f}
\end{eqnarray}
For later convenience, the flavor indices are numbered as $-1,0,1$. 
Under the $Z_3$ transformation (\ref{Z3_trans}), the FTBC (\ref{FTBC}) is changed into 
\begin{eqnarray}
q_f(x_4=\beta,{\vx})=-e^{-i\theta_f^\prime }q_f(x_4=0,{\vx}) 
\label{FTBC_change}
\end{eqnarray}
with 
\begin{eqnarray}
\theta_f^\prime ={2\pi\over{3}}(f-k)~~~~~(f=-1,0,1). 
\label{theta_f_new}
\end{eqnarray}
The boundary condition (\ref{FTBC_change}) after $Z_3$-transformation returns to the original one (\ref{FTBC}) by relabeling the flavor indices $f-k$ as $f$. 
Hence, if we consider  the FTBC instead of the standard one in $S$, 
the QCD-like theory obviously has $Z_3$ symmetry. This theory is 
shortly referred to as $Z_3$-QCD in this paper. 

The $Z_3$-QCD tends to QCD with symmetric three flavor quarks 
in the limit of $T\to 0$, since the difference between the two theories 
comes only from the temporal fermion boundary condition that 
has no contribution in the limit. 

When the fermion fields $q_f$ are transformed as~\cite{RW}
\begin{eqnarray}
q_f \to \exp{(-i\theta_fT\tau )}q_f ,
\label{transform_q}
\end{eqnarray}
the boundary condition (\ref{FTBC}) returns to the ordinary one
(\ref{boundary_q_u}), but ${\cal L}$ is changed into
\bea
{\cal L}^{\theta}= 
\bar{q}(\gamma_\nu D_\nu^\theta + m){q}
+{1\over{4g^2}}{F_{\mu\nu}^{a}}^2
\label{L_FTBC}
\eea
with $D_\nu^\theta \equiv \partial_\nu-i(A_\nu +\hat{\theta}
\delta_{\nu,4}T )$, where the flavor-dependent imaginary chemical potentials 
$i\hat{\theta}T$ are defined by
\begin{eqnarray}
\hat{\theta} &=& {\rm diag}(\theta_{-1},\theta_0,\theta_1)
\nonumber\\
&=& {\rm diag}(-2\pi/3,0,2\pi/3). 
\label{FTBC_chemical_p}
\end{eqnarray}
The flavor-dependent imaginary chemical potentials partially 
break $SU(3)$ flavor symmetry and associated $SU(3)$ chiral symmetry~\cite{Kouno_TBC,Sakai_TBC}.   
In the chiral limit $m\to 0$, global $SU_{\rm V}(3)\times SU_{\rm A}(3)$ symmetry
is broken down to  $(U(1)_{\rm V})^2\otimes (U(1)_{\rm A})^2$
\cite{Kouno_adjoint}. 
The symmetry is even broken into $(U(1)_{\rm V})^2$, as soon as chiral
symmetry is spontaneously broken.

\section{Sign problem and charge conjugation of gauge fields}
\label{real}

The grand canonical partition function of QCD with finite $\mu$ is obtained 
by 
\begin{eqnarray}
Z(\mu )=\int {\cal D}A_\mu {\cal D}\bar{q}{\cal D}q
e^{-S} 
\label{eq:Z_original}
\end{eqnarray}
with the action $S$ of Eqs.~(\ref{action})$\sim$(\ref{Anu}) in the previous section. The path integration over the quark fields leads to 
\begin{eqnarray}
Z(\mu )=\int {\cal D}A_\mu \det{[{\cal M} (\mu,A_\mu)]}e^{-S_{\rm G}}.
\label{eq:Z}
\end{eqnarray}
We start with the general case that 
$\mu$ and $A_\mu^a$ are complex.  
Using the relation 
\begin{eqnarray}
{\cal M} (\mu,A_\mu)^\dagger &=&-D_\mu (A_\mu^\dagger) \gamma_\mu +m-\mu^*\gamma_4
\nonumber\\
&=&\gamma_5{\cal M} (-\mu^*,A_\mu^\dagger)\gamma_5, 
\label{eq:Delta_dagger}
\end{eqnarray}
one can obtain 
\begin{eqnarray}
&&\left\{ \det{[{\cal M} (\mu,A_\mu)]}\right\}^*
=\det[{{\cal M} (\mu,A_\mu)^\dagger}]
\nonumber\\
&&=\det{\left[\gamma_5{\cal M} (-\mu^*,A_\mu^\dagger)\gamma_5\right]}
\nonumber\\
&&=\det{[\gamma_5]}\det{[\gamma_5]}\det{[{\cal M} (-\mu^*,A_\mu^\dagger) ]}
\nonumber\\
&&=\det{[{\cal M} (-\mu^*,A_\mu^\dagger) ]}.  
\label{eq:det_Delta}
\end{eqnarray}
In general, the last form of (\ref{eq:det_Delta}) does not agree with $\det{[{\cal M}(\mu,A_\mu )]}$. This leads to the fact that 
$\det{[{\cal M}(\mu,A_\mu )]}$ is not real generally. 
Even in the case that $\mu$ and $A_\mu^a$ are real, Eq. (\ref{eq:det_Delta}) does not ensure that $\det{[{\cal M}(\mu,A_\mu )]}$ is real.  
If the integrand is not real in (\ref{eq:det_Delta}), we cannot interpret the integrand as a probability.  
This makes it  impossible to apply the importance sampling method to the QCD action for the case of finite real $\mu$. 
This is nothing but the famous sign problem.   

The sign problem is originated in the fact that the fermion determinant ${\rm det}[{\cal M} (\mu, A_\mu )]$ is complex. 
For real $\mu$ and $A_\mu^a$, however, 
it is always possible to make $Z(\mu )$ real by  averaging the 
gauge configurations partially, as shown below. 
Using the charge conjugation matrix $C=\gamma_2\gamma_4$, one can get 
\begin{eqnarray}
[(C^t)^{-1}{\cal M} (-\mu^*,A_\mu^\dagger)C^t]^t &=&D_\mu(-A_\mu^*)\gamma_\mu+m+\mu^*\gamma_4
\nonumber\\
&=&{\cal M} (\mu^*,-A_\mu^*), 
\label{eq:Delta_dagger_2}
\end{eqnarray}
and hence ~\cite{Tanizaki}  
\begin{eqnarray}
&&\left( \det{[{\cal M} (\mu,A_\mu)]} \right)^*
=\det{[{\cal M} (-\mu^*,A_\mu^\dagger)]}
\nonumber\\
&&=\det{(C^t)^{-1}}\det{[{\cal M} (-\mu^*,A_\mu^\dagger )]}\det{C^t}
\nonumber\\
&&=\det{[(C^t)^{-1}{\cal M} (-\mu^*,A_\mu^\dagger )C^t]^t}
\nonumber\\
&&=\det{[{\cal M} (\mu^*,-A_\mu^*)]}. 
\label{eq:Delta_dagger_3}
\end{eqnarray}

In QCD with real $\mu$ and $A_\mu^a$, the relation $A_\mu^*=A_\mu^t$ leads to 
\begin{eqnarray}
\left( \det{[{\cal M} (\mu,A_\mu)]} \right)^*
&=&\det{[{\cal M} (\mu,-A_\mu^t)]}, 
\label{eq:Delta_dagger_3_real}
\end{eqnarray}
where $(A_\mu^a T^a)^t=A_\mu^a T^a$ for $a=1,3,4,6,8$ and $(A_\mu^a T^a)^t=-A_\mu^a T^a$ for $a=2,5,7$. 
Noting the relation \eqref{eq:Delta_dagger_3_real}, 
we consider a gauge configuration ${A_\mu^a}^\prime$ satisfying 
\begin{eqnarray}
A_\mu^\prime =\sum_{a=1}^8 {A_\mu^a}^\prime T_a=-\left(\sum_{a=1}^8 A_\mu^a T_a\right)^t=-A_\mu^t, 
\label{conf_A}
\end{eqnarray}
where ${A_\mu^a}^\prime =-A_\mu^a$ for $a=1,3,4,6,8$ and ${A_\mu^a}^\prime =A_\mu^a$ for $a=2,5,7$. 
The gauge action $S_{\rm G}$ and the Haar measure are invariant under the transformation $A_\mu \to A_\mu^\prime =-A_\mu^t$, since this transformation is nothing but charge conjugation for the gauge field. 
In fact, after this transformation, $F_{\mu\nu}$ is transformed into 
\begin{eqnarray}
{F_{\mu\nu}}^\prime &=&-\partial_\mu ({A_\nu}^t)+\partial_\nu ({A_\mu}^t)-ig[({A_\mu}^t),({A_\nu}^t)]
\nonumber\\
&=&-[F_{\mu\nu}]^{\tilde{t}},  
\label{Fmunudash}
\end{eqnarray}
where $\tilde{t}$ denotes the transpose operation only on the color index. 
Hence, the gauge action $S_{\rm G}$ is invariant under this transformation. 

In QCD with real $\mu$, we then obtain 
\begin{eqnarray}
\left( \det{[{\cal M} (\mu,A_\mu)]} \right)^*
&=&\det{[{\cal M} (\mu,-A_\mu^t)]}
\nonumber\\ 
&=&\det{[{\cal M} (\mu,A_\mu^\prime )]}. 
\label{eq:Delta_dagger_5}
\end{eqnarray}
Averaging the integrand of $Z(\mu)$ over 
the two configurations $A_\mu$ and $A_\mu^\prime$, one can get 
\begin{eqnarray}
&&{1\over{2}}\left(\det{[{\cal M} (\mu,A_\mu )]}e^{-S_{\rm G}}+\det{[{\cal M} (\mu,A_\mu^\prime )]}e^{-{S_{\rm G}}^\prime} \right)
\nonumber\\
&&={1\over{2}}\left[\det{[{\cal M} (\mu,A_\mu )]}+\left(\det{[{\cal M} (\mu,A_\mu)]} \right)^*\right] e^{-S_{\rm G}}
\nonumber\\
&&={\rm Re}\left(\det{[{\cal M} (\mu,A_\mu)]}\right)e^{-S_{\rm G}}.  
\label{eq:realization}
\end{eqnarray}
This ensures that the integrand of $Z(\mu)$ becomes real. 

Now we consider $Z_3$-QCD. 
The theory tends to QCD with symmetric three flavor quarks 
in the $T=0$ limit, since the difference between the two theories 
comes only from the temporal fermion boundary condition that 
has no contribution in the limit. 
Equation (\ref{eq:Delta_dagger_3_real}) is modified by the 
charge conjugation matrix as   
\begin{eqnarray}
&&\prod_{f=-1,0,1}\left( \det{[{\cal M} (\mu+i2\pi f/3,A_\mu)]} \right)^*
\nonumber\\
&&=
\prod_{f=-1,0,1} \det{[{\cal M} (\mu-i2\pi f/3,-A_\mu^t)]} 
\nonumber\\
&&=\prod_{f=1,0,-1} \det{[{\cal M} (\mu+i2\pi f/3,-A_\mu^t)]}
\nonumber\\
&&=\prod_{f=-1,0,1} \det{[{\cal M} (\mu+i2\pi f/3,-A_\mu^t)]}.
\nonumber\\
\label{eq:Delta_dagger_3_Z3}
\end{eqnarray}
In $Z_3$-QCD, the integrand of the partition function thus becomes real 
after the averaging procedure mentioned above. 
Even if the integrand of the partition function becomes real, it 
does not mean that the sign problem is solvable. 
This will be discussed later in Sec.~\ref{Potts}.

\section{Sign problem in 3-d 3-state Potts model}
\label{Potts}

We first construct a $Z_3$-symmetric Potts model 
to investigate the interplay between the sign problem and $Z_3$ symmetry. 
The standard 3-d 3-state Potts model is not $Z_3$-symmetric and 
has a sign problem, as shown below. 
Here we use the 3-d 3-state Potts model presented in Ref.~\cite{Alford}. 
The partition function of the 3-d 3-state Potts model 
is defined by~\cite{Karsch,Alford}
\begin{eqnarray}
&&Z( \kappa ,h )=\int {\cal D}\Phi e^{-S}=\int {\cal D}\Phi e^{-S_0}e^{-S_1}; 
\label{Potts_Z}\\
&&S_0[\Phi_{\vix},\kappa ]=-\kappa \sum_{{\vix},{\bf i}}\delta_{\Phi_{\vix},\Phi_{{\vix}+{\vii}}}, 
\label{Potts_S0}\\
&&S_1[\Phi_{\vix},h]=-\sum_{\vix} \left\{ h_+\Phi_{\vix}+h_-\Phi_{\vix}^*\right\}, 
\label{Potts_S1}\\
&&h_+=\exp{(-\beta (M-\mu ))},~~~h_-=\exp{(-\beta (M+\mu ))}
\nonumber\\
\label{Potts_h}
\end{eqnarray}
for the parameter $\kappa$ and the unit vector ${\vi}$ in 3-d space, where $M$ and $\mu$ correspond to the mass and the 
chemical potential of heavy quark, respectively. 
In Eq. (\ref{Potts_Z}), $\Phi_{\vix}$ represents a $Z_3$ element 
on site ${\vx}$ and plays a role of the Polyakov-loop operator 
in the Potts model, as already mentioned in Sec. \ref{sec:intro}.  
The $\Phi_{\vix}$ can be averaged over $\vx$ as 
\bea
{\bar \Phi}=\frac{1}{V}\sum_{{\vix}}\Phi_{{\vix}}=\frac{1}{V}\sum_{{\vix}}\Phi_{{\rm R},{\vix}}+\frac{1}{V}\sum_{{\vix}}\Phi_{{\rm I},{\vix}}={\bar \Phi}_{\rm R}+i {\bar \Phi}_{\rm I} , 
\nonumber\\
\label{averaged-Polyakov-loop}
\eea
where $\Phi_{{\vix},{\rm R}}$ (${\bar \Phi}_{\rm R}$) and $\Phi_{{\vix},{\rm I}}$ (${\bar \Phi}_{\rm I}$) mean the real and imaginary parts of $\Phi_{\vix}$ (${\bar \Phi}$), respectively, and $V$ is the lattice volume. 
We can regard the expectation value $\langle {\bar \Phi} \rangle$ 
as a "Polyakov loop" in the Potts model, since $\langle {\bar \Phi} \rangle$ 
has a property similar to the Polyakov loop $\langle L ({\vx})\rangle$ in QCD~\cite{DeGrand}. 
Obviously, $S_{0}$ is $Z_3$-invariant, but $S_{1}$ is not. 
In addition, $S_1$ can be complex for finite $\mu$.  

In the Potts model, the sign problem is originated in the fact that 
the effective chemical potential is complex. 
If we denote the phase of $\Phi_{\vix}$ as $\phi_{\vix}$, Eq. (\ref{Potts_S1}) is rewritten as 
\bea
S_1[\Phi_{\vix},h]=-2e^{-\beta M}\sum_{\vix}|\Phi_{\vix}|\cosh{(\beta  {\tilde{\mu}}_{\vix})} 
\label{Potts_S1_rewrite}
\eea
with the complex effective chemical potential 
${\tilde{\mu}}_{\vix}=\mu +i\phi_{\vix}$. 
When ${\tilde{\mu}}_{\vix}$ is complex,  
it makes $\cosh{{\tilde{\mu}}_{\vix}}$ complex in $S_1$ and consequently 
induces the sign problem. 
In this sense, the entanglement between $\mu$ and $i\phi_{\vix}$ is an 
origin of the sign problem in the Potts model.  
When $\mu$ is pure imaginary, so is $\tilde{\mu}$: namely $S_1^*=S_1$. 
This make the model free from the sign problem. 

The Potts model is constructed by simplifying 
the heavy-dense model of QCD 
in the limit of $M \to \infty$ and $\mu \to \infty$ 
with $M-\mu$ fixed at a finite value. 
Therefore, the Potts model should be considered 
for large $M$ and $\mu$. 
The term having $h_-$ is occasionally neglected, but is retained 
in the present paper.  

In the Potts model, $Z_3$ elements are taken as three states at each site on a 3-d lattice. 
Considering the $Z_3$ elements  as a substitute for the Polyakov-loop operator $L({\vx})$ in QCD, one can discuss the deconfinement transition through the 3-d Potts model.  
The operator  $L({\vx})$ depends on both ${\vx}$ and $T$, but information on $T$ is eliminated in the Potts model as a result of the substitution. 
Furthermore, in the 3-d Potts model, there is no temporal direction and the average over spatial volume should be taken on the Polyakov-loop.  
However, the Potts model has a parameter $\kappa$ in $S_0$. 
It was found in Ref.~\cite{DeGrand} that $\kappa$ dependence of $\langle {\bar \Phi} \rangle$ in the Potts model is similar to $T$ dependence of $\langle L ({\vx} ) \rangle$ in QCD. 
Although the relation between $\kappa$ and $T$ is not simple,  
we assume that larger  (smaller) $\kappa$ in the Potts model corresponds to higher (lower) $T$ in QCD and $\beta$ is just a parameter having the same dimension as $M$ and $\mu$. 
Below, we mainly use dimensionless parameters $\hat{M}=\beta M$ and $\hat{\mu}=\beta \mu$ instead of dimensionful ones $M$ and $\mu$.  

The factor $e^{-S_1}$ in (\ref{Potts_Z}) can be rewritten as $e^{-iS_{\rm I}-S_{\rm R}}$ with 
\begin{eqnarray}
S_{\rm R}[\Phi_{\vix}, \hat{M},\hat{\mu} ] &=&-2e^{-\hat{M}}\cosh{(\hat{\mu})}V{\bar \Phi}_{\rm R}, 
\label{S_R}\\
S_{\rm I}[\Phi_{\vix}, \hat{M},\hat{\mu} ] &=& -2e^{-\hat{M}}\sinh{(\hat{\mu})}V{\bar \Phi}_{\rm I}. 
\label{S_I}
\end{eqnarray}
Since $\Phi_{\vix}$ takes any value of $Z_3$ elements, $e^{i2\pi k/3}~(k=-1,0,1)$, $S_1$ becomes complex in general 
when $\mu \neq 0$. 
Hence, we take the configuration average over $\Phi_{\vix}$ and $\Phi_{\vix}^\prime =\Phi_{\vix}^*$, following the discussion in the previous section:
\begin{eqnarray}
&&{1\over{2}}\Bigl\{ e^{\left( -S_0[\Phi_{\vix},\kappa ]\right)}e^{\left( -S_1[\Phi_{\vix},h ] \right)}
+
e{\left(-S_0[\Phi_{\vix}^\prime ,\kappa ]\right)}e^{\left( -S_1[\Phi_{\vix}^\prime ,h ] \right)}\Bigr\}
\nonumber\\
&&=\cos{\left(S_{\rm I}[\Phi_{{\vix}, {\rm I}},\hat{M},\hat{\mu}]\right)}e^{\left( -S_0[\Phi_{\vix},\kappa ]-S_{\rm R}[\Phi_{{\vix},{\rm R}}, \hat{M}, \hat{\mu} ]\right)}
\label{Potts_average}
\end{eqnarray}
The integrand of the partition function (\ref{Potts_Z}) thus becomes real, 
but the cosine function becomes negative when 
\begin{eqnarray}
{\pi\over{2}}\le 2e^{-\hat{M}}\sinh{(\hat{\mu})}V\Phi_{\rm I}\le {3\pi\over{2}}~~~({\rm mod}[2\pi ]). 
\label{n_condition}
\end{eqnarray}
Hence, the sign problem appears even after the averaging procedure and it 
becomes stronger as $V$ increases. 

It is known that, although the Potts model has a sign problem in its path integral formulation, the model can be transformed into a "flux model" that has no complex action problem~\cite{Condella,Z3C}. 
However, it is not clear that such a transformation is possible or not in the case of  full LQCD simulations. 
Since we are interested in the sign problem itself, we treat the 3-d Potts model in the formulation where the sign problem is obvious. 

Since the 3-d 3-state Potts model has a sign problem in our formulation, namely, $F(\Phi_x)=e^{-S}$ becomes complex in general. 
Hence, we consider the following phase quenched approximation:  
\begin{eqnarray}
\langle O \rangle^\prime &=&{\int {\cal D}\Phi O (\Phi_{\vix} )F^\prime (\Phi_{\vix} )\over{Z^\prime}}; 
\label{PQ_Potts}\\
F^\prime (\Phi_{\vix} )&=&e^{-S_0-{S}_{\rm R}}, 
Z^\prime =\int {\cal D}\Phi F^\prime (\Phi_{\vix} ). 
\label{PQ_Potts_Z}
\end{eqnarray}
The phase factor $\langle F/F' \rangle$ and the true average $\langle O \rangle$ are then given by
\begin{eqnarray}
\langle {F\over{F^\prime}} \rangle^\prime &=&\langle e^{-iS_{\rm I}} \rangle^\prime ={\int {\cal D}\Phi 
[e^{-iS_{\rm I}}]F^\prime (\Phi_{\vix} )\over{Z^\prime}}
\nonumber\\
&=&{\int {\cal D}\Phi 
[\cos{S_{\rm I}}]F^\prime (\Phi_{\vix} )\over{Z^\prime}}=\langle \cos{S_{\rm I}} \rangle^\prime
={Z\over{Z^\prime}},   
\label{PQ_PF_Potts}\\
\langle O \rangle &=&{\int {\cal D}\Phi 
[O]F (\Phi_{\vix} )\over{Z}}
={\langle  O(\Phi_{\vix} )e^{-iS_{\rm I}}\rangle^\prime \over{\langle {F\over{F^\prime}} \rangle^\prime }}. 
\label{PQ_RW_Potts}
\end{eqnarray}

Since $\Phi_\vix$ can be complex in general, the calculation of the expectation value of the Polyakov-loop itself is somewhat complicated when the sign problem appears.  
Hence, for convenience, we calculate the absolute value 
$|{\bar \Phi} |$ instead of ${\bar \Phi}$ itself by using Monte Carlo simulations. 
To calculate physical quantities in a wide region of $\kappa$-$\mu$ plane, we took a rather small lattice with  spatial volume $V=6^3$. 
For this reason, in this paper we postpone the determination of the order of the deconfinement transition and simply define the transition point with 
the point where $\langle \bar{\Phi}\rangle \sim 0.5$.  
We also put $\hat{M}=10$ below.

\begin{figure}[htbp]
\begin{center}
\includegraphics[width=0.5\textwidth]{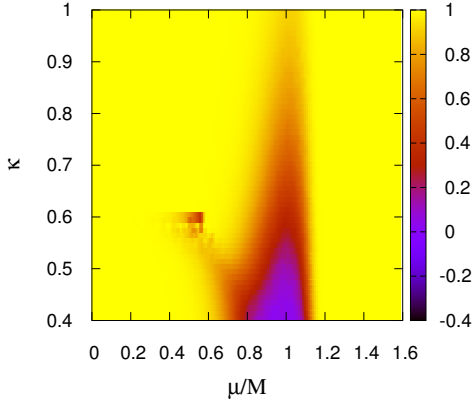}
\end{center}
\caption{The phase factor $\langle \cos{S_{\rm I}}\rangle ^\prime$ calculated with the phase quenched approximation in ordinary 3-d 3-state Potts model as a function of $\kappa$ and $\mu /M$. }
\label{Fig_ordinary_factor}
\end{figure}

Figure~\ref{Fig_ordinary_factor} shows the results of 
the phase factor  $\langle \cos{S_{\rm I}} \rangle^\prime$  
calculated with the quenched approximation 
in the $\kappa$-$\mu$ plane. 
The sign problem is serious 
in a ``triangle" area composed of three 
points $(\mu/M,\kappa) \approx (0.8,0.4),  (1.1,0.4), (1, 0.6)$, 
since $\langle \cos{S_{\rm I}} \rangle^\prime$ is small or negative 
there. 

\begin{figure}[htbp]
\begin{center}
\includegraphics[width=0.5\textwidth]{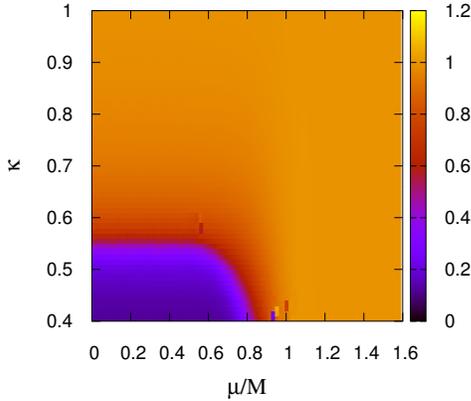}
\end{center}
\caption{The expectation value $\langle |{\bar \Phi} |\rangle$ calculated with the reweighting method in ordinary 3-d 
3-state Potts model as a function of $\kappa$ and $\mu /M$.     
}
\label{Fig_ordinary_diag}
\end{figure}

Figure~\ref{Fig_ordinary_diag} shows the results of 
$\langle |{\bar \Phi} |\rangle$ 
calculated with the reweighting method  
in the $\kappa$-$\mu$ plane. 
The line where $\langle |{\bar \Phi} |\rangle$ 
rapidly changes can be regarded as a ``critical line" 
$\kappa=\kappa_{c}(\mu/M)$.   
The value of $\kappa_{c}(\mu/M)$ is about 0.55 for $\mu/M <0.5$, but rapidly 
decreases from 0.55 to 0.4 as $\mu/M$ increases from 0.5 to 0.75. 

Comparing Fig.~\ref{Fig_ordinary_factor} with Fig.~\ref{Fig_ordinary_diag}, 
we can see, as an interesting property, that 
$\langle \cos{S_{\rm I}} \rangle^\prime$ tends to be  
small on the critical line particularly in $0.4 < \mu/M <0.75$. 
In the triangle area composed of three 
points $(\mu/M,\kappa) \approx (0.8,0.4),  (1.1,0.4), (1, 0.6)$, the value of 
$\langle |{\bar \Phi} |\rangle $ is larger than 1 or negative, 
since $\langle \cos{S_{\rm I}} \rangle^\prime$ is small or 
negative.  
The unnatural behavior of $\langle |{\bar \Phi} |\rangle$
is caused by that of $\langle \cos{S_{\rm I}} \rangle^\prime$. 
Therefore, physical discussion cannot be made in the triangle region. 

\section{$Z_3$-symmetric 3-d Potts model}
\label{Z3Potts}

Taking the same way as the extension of QCD to $Z_3$-QCD, we now 
construct a $Z_3$-symmetric 3-d Potts model from 
the 3-d Potts model defined in the previous section. 
The action $S_{Z_3}$ of the new model is assumed to be described in 
a power series of $\Phi_{\vix}$ and $\Phi^*_{\vix}$ just as the original Potts model, 
and the flavor-dependent imaginary chemical potentials are 
introduced by 
replacing $h_+\Phi_{\vix}$, $h_-\Phi^*_{\vix}$ by 
\begin{eqnarray}
h_{+,f}\Phi_{\vix}=e^{i2\pi f/3}\exp{(\hat{\mu}-\hat{M})}\Phi_{\vix}, \label{hplus}\\
h_{-,f}\Phi^\ast_{\vix}
=e^{-i2\pi f/3}\exp{(-\hat{\mu}-\hat{M})}\Phi^\ast_{\vix}. \label{hminus}
\end{eqnarray}
Here we consider $Z_3$ elements as three states of $\Phi_{\vix}$ 
for a while, but will increase the number of states later in which 
some states do not belong to the group $Z_3$. 

The pure gauge and first-order terms, $S_{0,Z_3}$ and $S_{1,Z_3}$, 
of $S_{Z_3}$ are then defined as 
\begin{eqnarray}
S_{0,Z_3}[\Phi_{\vix},\kappa ]&=&-\kappa \sum_{{\vix},{\vii}}|\Phi_{\vix}||\Phi_{{\vix}+{\vii}}|\delta_{\Phi_{\vix},\Phi_{{\vix}+{\vii}}}, 
\label{Z3_Potts_S0}\\
S_{1,Z_3}[\Phi_{\vix}, h_{\pm}]&=&-\sum_{f=-1,0,1}\sum_{{\vix}}\left(h_{+,f}\Phi_{\vix} +h_{-,f}\Phi^*_{\vix}\right)
\label{Z3_Potts_1} .
\end{eqnarray}
Here, $S_{0,Z_3}$ is $Z_3$-invariant, but $S_{1,Z_3}$ is $Z_3$-variant. 
Note that $S_{0,Z_3}$ is reduced to $S_0$ when the values of $\Phi_{\vix}$ are 
restricted to $Z_3$ elements only.   
If the $f$ summation is taken in $S_{1,Z_3}$, the term vanishes as follows:    
\begin{eqnarray}
&&S_{1,Z_3}[\Phi_{\vix}, h_{\pm}]=-\sum_{f=-1,0,1}\sum_{{\vix}}\left(h_{+,f}\Phi_{\vix} +h_{-,f}\Phi^*_{\vix}\right) \nonumber\\
&&=\sum_{{\vix}}(h_+\Phi_{\vix} +h_-\Phi^*_{\vix})\sum_{f=-1,0,1}e^{i2\pi f/3} =0 .
\label{Z3_Potts_1-1}
\end{eqnarray}
Similarly, any $Z_3$-variant terms vanish after the $f$ summation. 
Considering the power series of $S_{Z_3}$ 
up to the third order of $\Phi_{\vix}$ and $\Phi_{\vix}^*$, 
we can construct 
a $Z_3$-symmetric Potts model as 
\begin{eqnarray}
&&Z( \kappa, h_M, h_{\pm}) 
=\int {\cal D}\Phi e^{-S_{Z_3}}
=\int {\cal D}\Phi e^{-S_{0,Z_3}}e^{-S_{2+3,Z_3}} , 
\nonumber\\
\label{Potts_Z3}
\end{eqnarray}
where $h_M=e^{-\hat{M}}$ and 
\begin{eqnarray}
S_{{2+3},Z_3}[\Phi_{\vix},h_M,h_{\pm}]&=&S_{2,Z_3}[\Phi_{\vix},h_M]+S_{3,Z_3}[\Phi_{\bf x},h_{\pm}];\nonumber\\
\label{Z3_Potts_2_3}\\
S_{2,Z_3}[\Phi_{\vix}, h_M]&=&-g_2{h_M}^2\Phi_{\vix}\Phi^*_{\vix}, \label{Z3_Potts_2}\\
S_{3,Z_3}[\Phi_{\vix}, h_{\pm}]&=&-g_3\left( {h_+}^3{\Phi_{\vix}}^3 +{h_-}^3{\Phi^*_{\vix}}^3 \right) 
\label{Z3_Potts_3}
\end{eqnarray}
with the coupling parameters $g_2$ and $g_3$. 
For simplicity, we take $g_2=g_3=1$. 
Note that, e.g., the third-order terms of $\Phi_x$ can appear  when the logarithmic Fermionic effective Lagrangian~\cite{Greensite:2014cxa} 
is expanded by $\Phi_\vix$ as
\begin{eqnarray}
&&\sum_{f=-1,0,1}\log{\left(1+3h_{+,f}\Phi_\vix +3h_{+,f}^2\Phi_\vix^* +(h_{+,f})^3\right)}
\nonumber\\
&&+\sum_{f=-1,0,1}\log{\left(1+3h_{-,f}\Phi_\vix^* +3h_{-,f}^2\Phi_\vix +(h_{-,f})^3\right)}
\nonumber\\
&&=\log{\left[\prod_{f=-1,0,1}\left(1+3h_{+,f}\Phi_\vix+3h_{+,f}^2\Phi_\vix^*+(h_{+,f})^3\right) \right]}
\nonumber\\
&&+\log{\left[\prod_{f=-1,0,1}\left(1+3h_{-,f}\Phi_\vix^*+3h_{-,f}^2\Phi_\vix+(h_{-,f})^3\right) \right]}
\nonumber\\
&&=\log{\left[\left( a+27(h_+)^3\Phi_\vix^3+27(h_+)^6{\Phi_\vix^*}^3+\cdots \right) \right]}
\nonumber\\ 
&&+\log{\left[\left( b+27(h_-)^3{\Phi_\vix^*}^3+27(h_-)^6{\Phi_\vix}^3+\cdots \right) \right]} ,
\label{log_action}   
\end{eqnarray}
 where  
$a$ and $b$ are $\Phi_\vix$-independent real quantities.
On the other hand, the terms proportional to ${\Phi_\vix}{\Phi_\vix}^*$ is expected to appear if the mesonic contribution in the effective Lagrangian is expanded. 

The factor $e^{-S_{2+3,Z_3}}$ can be rewritten into 
\begin{eqnarray}
&&e^{-S_{2,Z_3} -S_{3,Z_3}}=e^{-iS_{{\rm I},Z_3}}e^{-S_{{\rm R},Z_3}}e^{-S_{2,Z_3}}
\label{Z3_Potts_osc}
\end{eqnarray}
with 
\begin{eqnarray}
S_{{\rm R},Z_3}[\Phi_{\vix},\hat{M},\hat{\mu}]&=&-2e^{-3\hat{M}}\cosh{(3\hat{\mu})}\sum_{{\vix}}({\Phi_{\vix}}^3)_{\rm R} , \nonumber\\
\label{SR_Z3}\\ 
S_{{\rm I},Z_3}[\Phi_{\vix},\hat{M},\hat{\mu}]&=&-2e^{-3\hat{M}}\sinh{(3\hat{\mu})}\sum_{{\vix}}({\Phi_{\vix}}^3)_{\rm I} \nonumber\\ 
\label{SI_Z3}
\end{eqnarray}
for the real and imaginary parts, 
$({\Phi_{\vix}}^3)_{\rm R}$ and $({\Phi_{\vix}}^3)_{\rm I}$, 
of ${\Phi_{\vix}}^3$.  
When the values of $\Phi_{\vix}$  are restricted to $Z_3$ elements, 
both $S_{0,Z_3}$ and $S_{2+3,Z_3}$ are real, 
indicating that no sign problem takes place.
This model is referred to as 
``$Z_3$-symmetric 3-state Potts model" in the present paper. 

\begin{figure}[htbp]
\begin{center}
\includegraphics[width=0.5\textwidth]{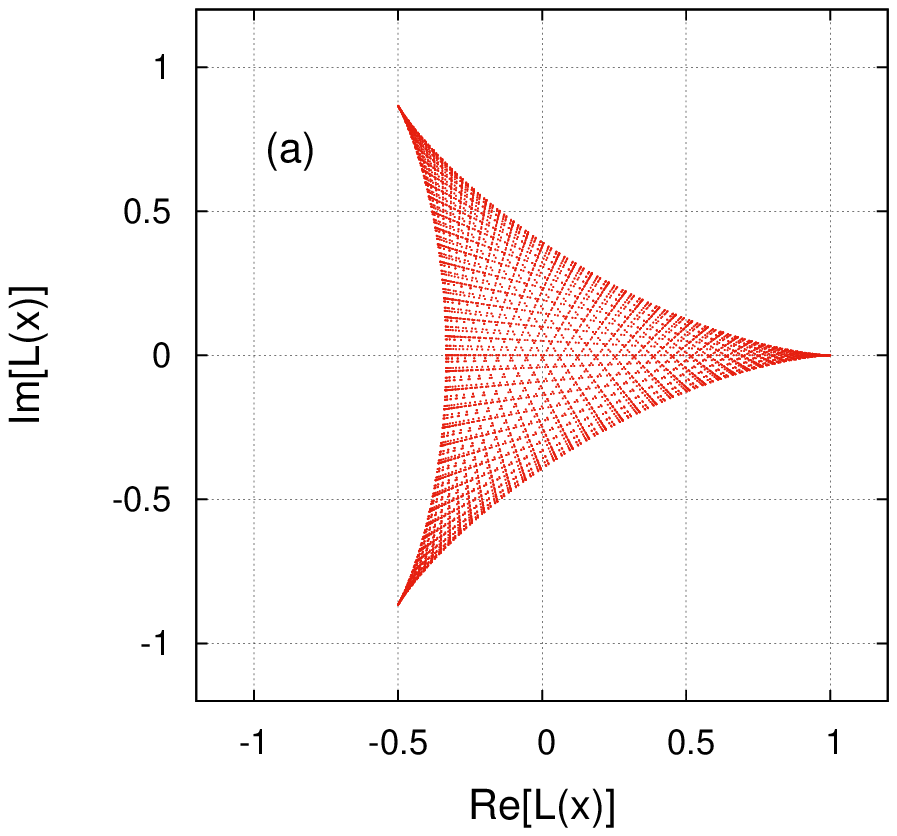}
\includegraphics[width=0.5\textwidth]{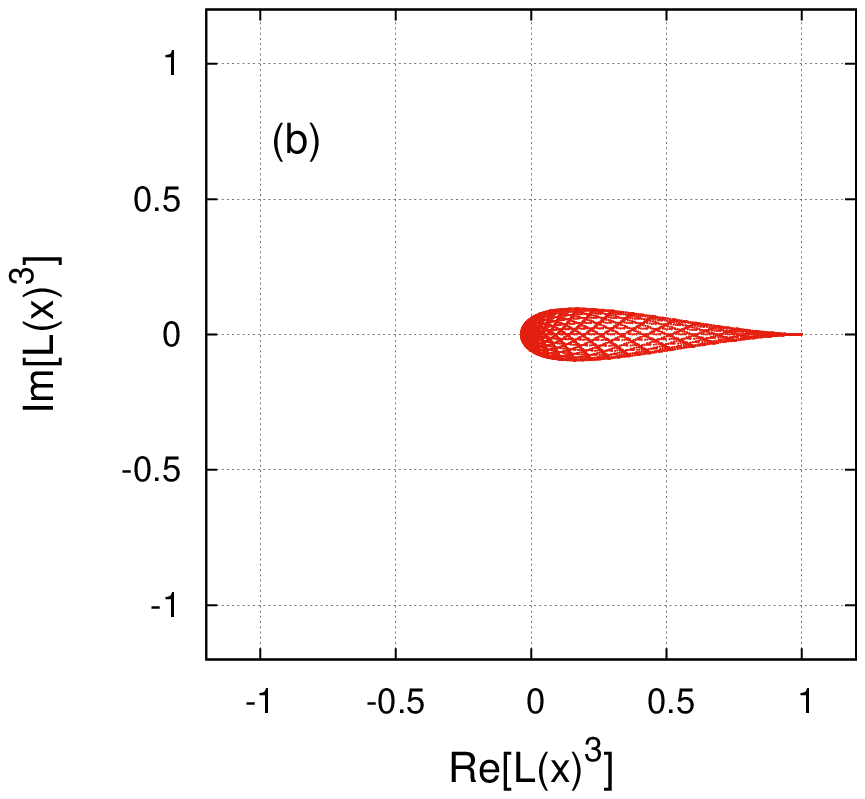}
\end{center}
\caption{Allowed regions of (a) $L({\vx})$ and (b) $L^3({\vx})$ in the complex plane.}
\label{Fig_Phi_QCD}
\end{figure}

Meanwhile, the $Z_3$-symmetric heavy-dense model of QCD has the sign problem,  since the Polyakov-loop operator $L({\vx})$ can take not only $Z_3$ elements but also ones not belonging to  the group $Z_3$. 
In the model, $L({\vx})$ can be parameterized as~\cite{Greensite:2014cxa}
\begin{eqnarray}
L({\vx})={1\over{3}}\left( e^{i\phi_1({\vix})}+e^{i\phi_2({\vix})}+e^{-i(\phi_1({\vix})+\phi_2({\vix}))}\right) , 
\label{Phi_para}
\end{eqnarray}
and the region that $L({\vx})$ can take is illustrated 
in Fig.~\ref{Fig_Phi_QCD}(a). 
The region has a hyperbolic-triangle-like shape.  
Three vertices of the triangle correspond to $Z_3$ elements, 
and the other points in the region do not belong to the group $Z_3$. 
Figure~\ref{Fig_Phi_QCD}(b) corresponds to the region that $L^3({\vx})$ is allowed to take. 
The allowed region of $L^3({\vx})$ is well localized near the real axis compared with that of $L({\vx})$. 
 This suggests that  the sign problem is less serious in $Z_3$-QCD than in QCD~\cite{Kouno_DFTBC}.  
Since it is not easy to confirm this suggestion directly in QCD, we will 
compare three types of models: 
(i) the original  Potts model, 
(ii) $Z_3$-symmetric 3-state 
Potts model defined above, 
and 
(iii) $Z_3$-symmetric {\it several-state} Potts models, 
each with different number of states larger than 3. 
$Z_3$-symmetric several-state Potts models 
is an extension of $Z_3$-symmetric 3-state 
Potts model and 
is not free from the sign problem in general.

Before constructing 
$Z_3$-symmetric several-state Potts models explicitly, 
we make the following preparation. 
When the $\Phi_{\vix}$ are allowed to take values not belonging 
to the group $Z_3$  just as the heavy quark model, the term $S_{2+3,Z_3}$ can be complex. 
In that case, we take the average of $\Phi_{\vix}$ and $\Phi^\prime_{\vix} =\Phi^*_{\vix}$ configurations to make the partition function real, 
following the discussion in the previous section. 
This leads to  
\begin{eqnarray}
&&{e^{-S_{0,Z_3}[\Phi_{\vix}]}e^{ -S_{2+3,Z_3}[\Phi_{\vix}]}
+
e^{-S_{0,Z_3}[\Phi_{\vix}^\prime ]}e^{-S_{2+3,Z_3}[\Phi_{\vix}^\prime ]}\over{2}}
\nonumber\\
&&=\cos{\left( S_{{\rm I},{\rm },Z_3}[\Phi_{\vix}]\right)}
e^{\Bigl( -S_{0,Z_3}[\Phi_{\vix}]-S_{2,Z_3}[\Phi_{\vix}]-S_{{\rm R},Z_3}[\Phi_{\vix}]\Bigr)} . 
\nonumber\\
\label{Z3_Potts_average}
\end{eqnarray}
The integrand of the partition function (\ref{Potts_Z3}) thus becomes real, 
but the cosine function becomes negative when 
\begin{eqnarray}
{\pi\over{2}}\le 2e^{-3\hat{M}}\sinh{(3\hat{\mu})} \sum_{{\vix}}\left({\Phi_{\vix}}^3\right)_{\rm I}\le {3\pi\over{2}}~~~({\rm mod}[2\pi ]). 
\nonumber\\ 
\label{n_condition_Z3}
\end{eqnarray}
As mentioned in Fig.~\ref{Fig_Phi_QCD}, however, 
the absolute value of imaginary part ${\rm Im} [L({\vx})^3]$  is much smaller than that of ${\rm Im} [L({\vx})]$. 
This is true also in $Z_3$-symmetric Potts models  with several states; 
namely, ${\rm max}[|({\Phi^3_{\vix}})_{\rm I}|] \ll {\rm max}[|(\Phi_{\vix})_{\rm I}|]$. 
Hence, it can be expected that the sign problem is less serious 
in $Z_3$-symmetric Potts model with several states 
than in the original Potts model without exact $Z_3$ symmetry.

Now  we explicitly consider four kinds of $Z_3$-symmetric Potts models, 
each with 3, 4, 7 and 13 states: 

\begin{enumerate}
\renewcommand{\labelenumi}{(\Alph{enumi})} 
\item $Z_3$-symmetric 3-d 3-state Potts model with 
\begin{eqnarray}
\{ \Phi_{\vix} \} =\{1,e^{\pm i2\pi/3}\}, 
\label{3-state}
\end{eqnarray}
which are denoted by three solid triangle symbols in Fig.~\ref{various_Z3}.  
This model is free from the sign problem, since 
$S_{{2+3},Z_3}$ is always real.  

\begin{figure}[htbp]
\begin{center}
\includegraphics[width=0.5\textwidth]{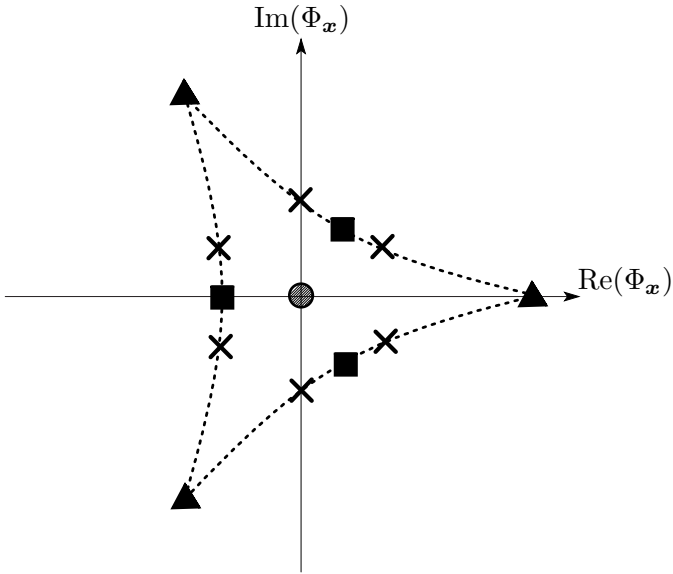}
\end{center}
\caption{ Values of $\Phi_x$ taken in $Z_3$-symmetric 3-d Potts models (A)$\sim$(D).}
\label{various_Z3}
\end{figure}

\item $Z_3$-symmetric 3-d 4-state Potts model with 
\begin{eqnarray}
\{ \Phi_{\vix} \} =\{1,e^{\pm i2\pi/3},0\}, 
\label{4-state}
\end{eqnarray}
which are denoted by three solid triangles and a solid circle 
in Fig.~\ref{various_Z3}. 
The sign problem does not appear. 

\item $Z_3$-symmetric 3-d 7-state Potts model with
\begin{eqnarray}
\{ \Phi_{\vix} \}=\{1,e^{\pm i2\pi/3},0,-1/3,e^{\pm i\pi /3}/3 \}, 
\label{7-state}
\end{eqnarray}
which are denoted by three solid triangles, a solid circle and 
three solid squares in Fig.~\ref{various_Z3}. 
The sign problem does not appear. 

\item $Z_3$-symmetric 3-d 13-state Potts model with 
\begin{eqnarray}
\{ \Phi_{\vix} \} =\{1,e^{\pm i2\pi/3},0,-1/3,e^{\pm i\pi /3}/3,
\nonumber\\
re^{\pm i\pi /6},re^{\pm i\pi /2},re^{\pm i5\pi /6} \} , 
\label{13-state}
\end{eqnarray}
where $r$ is taken to be 0.4. 
Equation (\ref{13-state}) is denoted by three solid triangles, a solid circle, three solid squares 
and six crosses in Fig.~\ref{various_Z3}. 
Note that, except for the origin, the other twelve values are chosen to lie on the boundary of the region where the value of $L({\vx})$ can be taken in Fig.~\ref{Fig_Phi_QCD} (a). 
The sign problem comes from six crosses 
$\{ \Phi_{\vix}  \}=
\{r e^{\pm i\pi /6},re^{\pm i\pi /2},re^{\pm i5\pi /6}\}$, 
since $(\Phi_{\vix})^3$ has a finite phase in this case. 
\end{enumerate}

Properties of the four $Z_3$-symmetric Potts models (A)-(D) 
are summarized in Table \ref{tb:models}, together with those of 
the original 3-d 3-state Potts model. 

Before closing this section, we add one more comment. 
If we denote $\Phi_\vix=r_\vix\exp{(i\phi_\vix)}$, where $r_\vix$ and $\phi_\vix$ are the absolute value and the phase of $\Phi_\vix$, respectively,   
$(\Phi_\vix)^3=r_\vix^3\exp{(3i\phi_\vix)}=r_\vix^3(\cos{(3\phi_\vix)}+i\sin{(3\phi_\vix)})$.  
Therefore, if $r_\vix$ is small, the sign problem is not severe. 
Hence, it can be expected that, for fixed $\phi_\vix$, larger $r_\vix$ configurations dominates the sign problem. 
So we choose our simulation points on the boundary of the hyperbolic-triangle in Fig.~\ref{various_Z3}, except for the origin, the special point for confinement. 
Furthermore, to keep the $Z_3$ symmetry, we use the configurations of $\Phi_\vix$ with the discrete phase $\phi_\vix^j =2\pi j/(3n)~(j=0,1,2,\cdots n-1)$, where $n$ is a positive integer. 
In principle, the calculation is possible by using the other points on the boundary with larger $n>4$, but it requires huge computing time. 
We postpone estimation of the contributions of other points is open question in future.

\section{Sign problem in $Z_3$-symmetric Potts model}
\label{Numerical}

In this section, we perform numerical simulations for 
$Z_3$-symmetric Potts models (A)-(D) and compare the results with 
those of the original Potts model. 
In particular, the expectation value $\langle |{\bar \Phi}| \rangle$ 
of the absolute value $|{\bar \Phi}|$   
is calculated to determine the confinement-deconfinement transition line. 
The purposes of this analyses  are to  (1) to clarify the interplay between $Z_3$ symmetry and the sign problem, and 
(2) to see how the deconfinement transition line changes in $\kappa$-$\mu$ plane when the value of $\Phi_{\vix}$ is extended bit by bit. 
As mentioned in Sec.~\ref{Potts}, the standard Metropolis algorithm 
is taken for the simulations.

\subsection{$Z_3$-symmetric 3-d 3-state Potts model}

\begin{figure}[htbp]
\begin{center}
\includegraphics[width=0.5\textwidth]{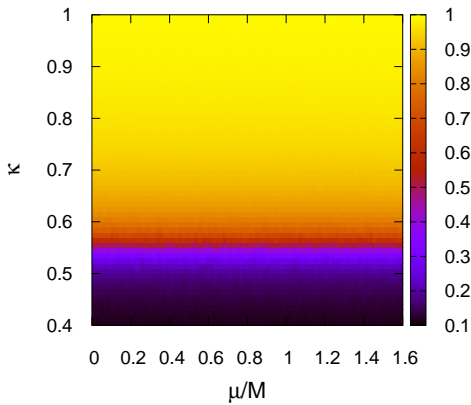}
\end{center}
\caption{The expectation value $\langle{|{\bar \Phi}|}\rangle$ in $Z_3$-symmetric 3-d 
3-state Potts model as a function of $\kappa$ and $\mu/M$.    
}
\label{Fig_Z3_3s}
\end{figure}

\begin{figure}[htbp]
\begin{center}
\includegraphics[width=0.5\textwidth]{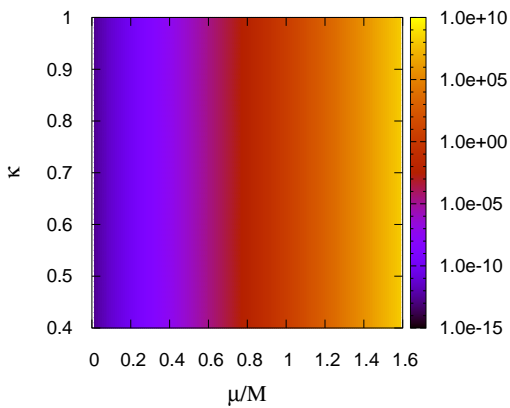}
\end{center}
\caption{The expectation value $n$ in $Z_3$-symmetric 3-d 3-state Potts model as a function of $\kappa$ and $\mu/M$.
Note that $n$ is dimensionless, since the dimensionless volume $V=6^3$ is used in calculations of (\ref{n_density}).
}
\label{Fig_n_density}
\end{figure}

Figure~\ref{Fig_Z3_3s} shows the expectation value $\langle |{\bar \Phi}| \rangle$ in the $\kappa$-$\mu$ plane for model (A). Note that the reweighting is not necessary for this model, since the model has no sign problem. 
There appears a rapid change of $\langle |{\bar \Phi}| \rangle$ at $\kappa =0.55$, independently of $\mu$.  
The $\mu$ independence comes from the fact that the $\mu$-dependent term $S_{2+3,Z_3}$ becomes constant 
because of $\Phi_{\vix}(\Phi_{\vix})^*={\Phi_{\vix}}^3={\Phi^*_{\vix}}^3=1$. 
In fact, $\langle |{\bar \Phi}| \rangle$ can be rewritten as 
\begin{eqnarray}
\langle |{\bar \Phi} | \rangle &=&
{\int {\cal D}\Phi |\Phi |e^{-S_{0,Z_3}-S_{2+3,Z_3}}\over{\int {\cal D}\Phi e^{-S_{0,Z_3}-S_{2+3,Z_3}}}}
\nonumber\\
&=&
{e^{-S_{2+3,Z_3}}\int {\cal D}\Phi |\Phi |e^{-S_{0,Z_3}}\over{e^{-S_{2+3,Z_3}}\int {\cal D}\Phi e^{-S_{0,Z_3}}}}
\nonumber\\
&=&
{\int {\cal D}\Phi |\Phi |e^{-S_{0,Z_3}}\over{\int {\cal D}\Phi e^{-S_{0,Z_3}}}}. 
\label{Phi_averaged}
\end{eqnarray}
The final form of Eq. (\ref{Phi_averaged}) has no $\mu$-dependence.    
However, this does not means that all quantities do not depend on $\mu$.  
In fact, the number density 
\begin{eqnarray}
n &=&{1\over{\beta V}}{\partial \left(\log{Z}\right)\over{\partial \mu}}={1\over{V}}{\partial \left(\log{Z}\right)\over{\partial \hat{\mu}}}
\nonumber\\
&=&{6e^{-3\hat{M}}\over{V}}\Bigl[\sinh{(3\hat{\mu})}\langle  \sum_{\vix} (\Phi_x^3)_{\rm R} \rangle 
\nonumber\\
&&+i\cosh{(3\hat{\mu})}\langle \sum_{\vix} (\Phi_{\vix^3})_{\rm I} \rangle\Bigr]
\label{n_density}
\end{eqnarray}
has $\mu$-dependence, as shown in Fig.~\ref{Fig_n_density}. 
(Note that (\ref{Phi_averaged}) is true only in the 3-d 3 state $Z_3$-symmetric Potts model, whereas (\ref{n_density}) is true for all $Z_3$-symmetric Potts model. )   
The quark number density $n$ can be regarded as an order parameter for the onset phase transition that is located at $\mu/M =1$ and $T=0$.       
Below we will see how the two transitions entangle each other and the sign problem appears when the value region 
of $\Phi_{\vix}$ is extended bit by bit. 
Note that the second term in the second line of (\ref{n_density}) vanishes when $(\Phi_{\vix}^3)_{\rm I}=0$ and the sign problem does not appear.

\begin{figure}[htbp]
\begin{center}
\includegraphics[width=0.5\textwidth]{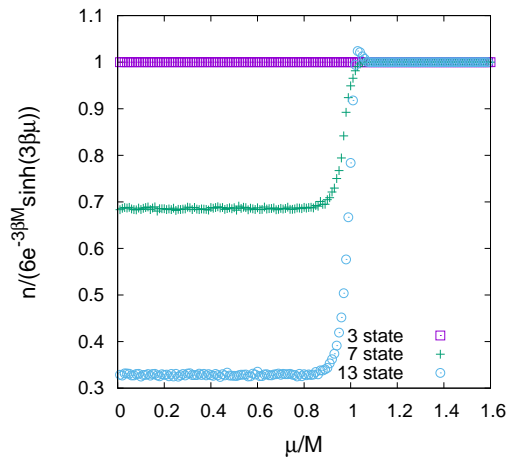}
\end{center}
\caption{$\mu$-dependence of $n$ at $\kappa =0.65$ in $Z_3$-symmetric 3-d Potts model. 
The boxes, the crosses and the circles represent the results in 3, 7 and 13-state Potts model. }
\label{Fig_n_density_kappa}
\end{figure}

In Fig.~\ref{Fig_n_density_kappa}, $\mu$-dependence of $n$ 
is shown at $\kappa =0.65$ for three $Z_3$-symmetric Potts models.  
When $\kappa =0.65$,  in $Z_3$-symmetric 3-state Potts model, the onset of $n$ is smooth and 
$n$ is proportional to $e^{-3\hat{M}}\sinh{3\hat{\mu}}$ in the model, since $(\Phi_{\vix}^3)_{\rm R}=1$ and $(\Phi_{\vix}^3)_{\rm I}=0$ in (\ref{n_density}). 

It is interesting that, in $Z_3$-symmetric 3-d 3-s Potts model, the resulting deconfinement transition line then becomes a horizontal line in the $\mu$-$\kappa$ plane. 
Similar property is also seen in $SU(N)$ gauge theory in the large $N$ limit~\cite{MacLerran,Hidaka}, where the contribution of the fermion with $N$ degrees of freedom to the thermodynamic potential is negligible compared with that of gauge field with $N^2-1$ degrees of freedom.  
In the case of $Z_3$-symmetric theory, $Z_3$ symmetry suppresses the fermion contribution and weakens the $\mu$-dependence of the critical temperature of deconfinement transition~\cite{Sakai_TBC} .

\subsection{$Z_3$-symmetric 3-d 4-state Potts model}

\begin{figure}[htbp]
\begin{center}
\includegraphics[width=0.5\textwidth]{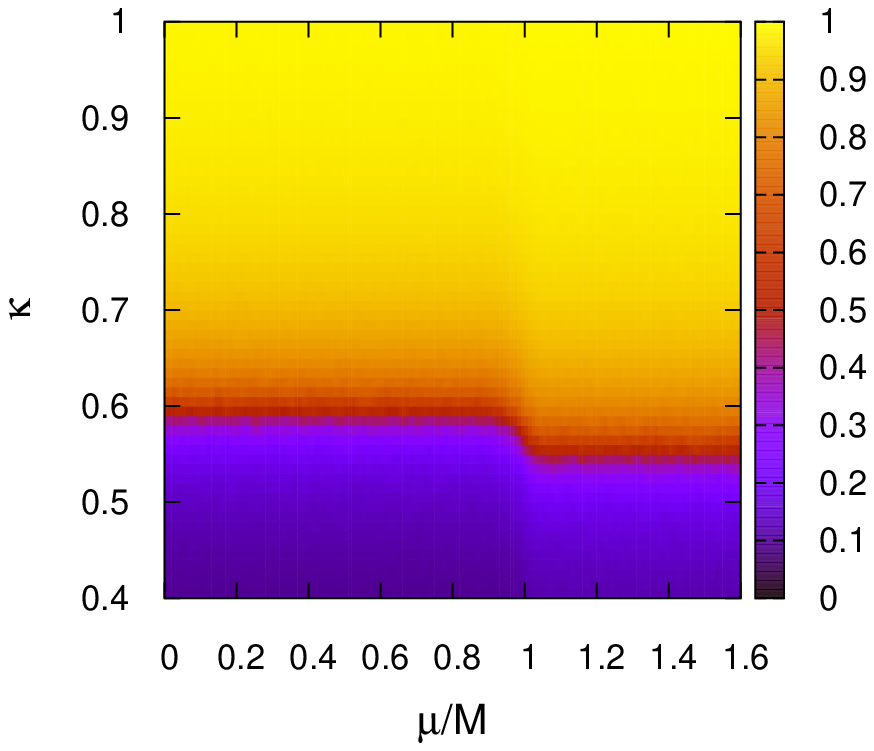}
\end{center}
\caption{The expectation value $\langle{|{\bar \Phi}|}\rangle$ in $Z_3$-symmetric 3-d 
4-state Potts model as a function of $\kappa$ and $\mu/M$. }
\label{Fig_Z3_4s}
\end{figure}

In this case, the model has no sign problem as is in the previous one. 
Figure~\ref{Fig_Z3_4s} shows the expectation value $\langle |{\bar \Phi}| \rangle$ in the $\kappa$-$\mu$ plane. 
The deconfinement transition line $\kappa=\kappa_{c}(\mu/M)$ defined by a rapid change of $\langle |{\bar \Phi}| \rangle$ is located at $\kappa =0.60$ for $\mu/M < 1$, but goes down to $\kappa =0.55$ at $\mu/M \approx 1$, and keeps $\kappa =0.55$ for $\mu/M > 1$. 

Thus, the deconfinement transition is beginning to have $\mu$-dependence. 

\subsection{$Z_3$-symmetric 3-d 7-state Potts model}

\begin{figure}[htbp]
\begin{center}
\includegraphics[width=0.5\textwidth]{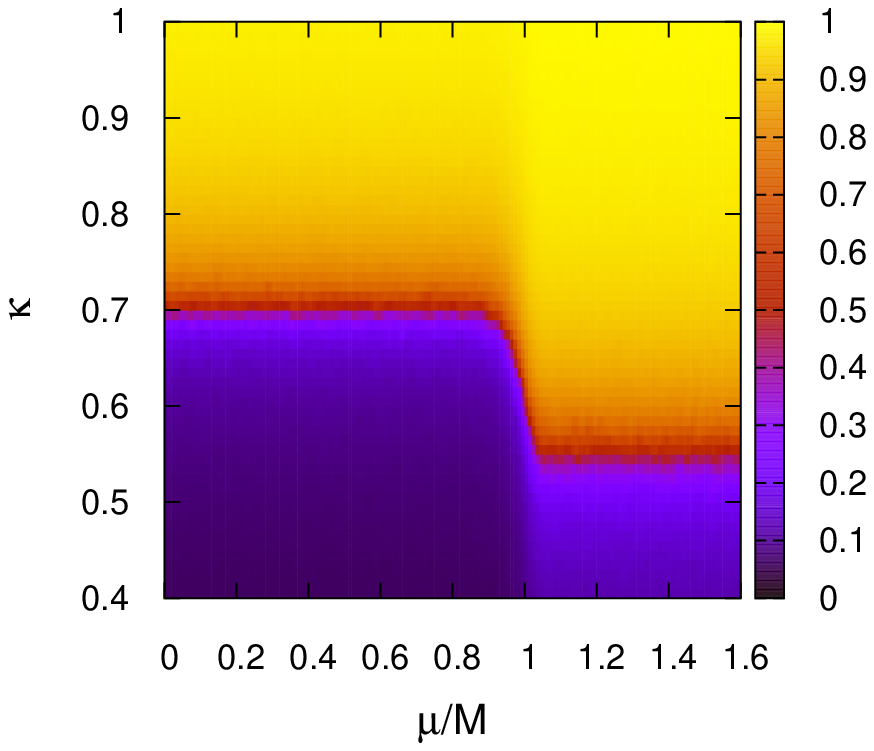}
\end{center}
\caption{The expectation value $\langle{|{\bar \Phi}|}\rangle$ in $Z_3$-symmetric 3-d 
7-state Potts model as a function of $\kappa$ and $\mu/M$.     
}
\label{Fig_Z3_7s}
\end{figure}

\begin{figure}[htbp]
\begin{center}
\includegraphics[width=0.5\textwidth]{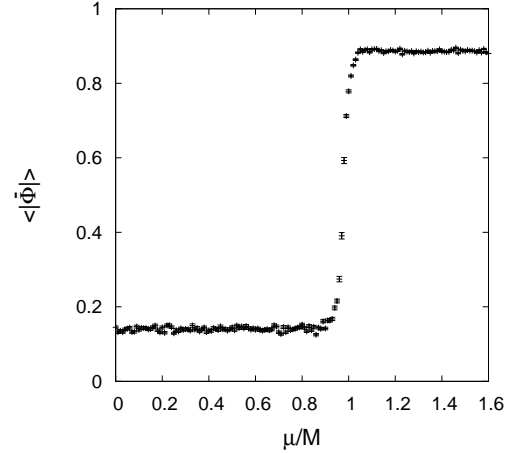}
\end{center}
\caption{$\mu$-dependence of $\langle|{\bar \Phi}|\rangle$ at $\kappa =0.65$ in $Z_3$-symmetric 3-d 7-state Potts model. }
\label{Fig_Z3_7s_Phi}
\end{figure}

In this case, the model has no sign problem as is in the previous two cases. 
Figure~\ref{Fig_Z3_7s} shows the expectation value  $\langle |{\bar \Phi}| \rangle$  in the $\kappa$-$\mu$ plane. 
The deconfinement transition line $\kappa=\kappa_{c}(\mu/M)$ is located at $\kappa =0.7$ for $\mu/M < 1$, but goes down to $\kappa =0.55$ at $\mu/M \approx 1$, and keeps $\kappa =0.55$ for $\mu/M > 1$. 
The deconfinement transition line thus has stronger $\mu$-dependence in model (C) than in model (B). 
Figure~\ref{Fig_Z3_7s_Phi} shows $\mu$-dependence of 
$\langle |{\bar \Phi}| \rangle$ with $\kappa$  fixed at 0.65.  
We can see that $\langle |{\bar \Phi}| \rangle$ suddenly increases at 
$\mu/M \approx 1$.  
In Fig.~\ref{Fig_n_density_kappa}, $\mu$-dependence of  $n$ is shown. 
In this model, $n$ also rapidly increases at $\mu/M \approx 1$.  

\subsection{$Z_3$-symmetric 3-d 13-state Potts model}

In model (D), when $S_{2+3,Z_3}$ beoomes complex,  
$F(\Phi_\vix )=e^{-S}$ also does, so that the sign problem appears. 
We then use the following phase quenched approximation: 
\begin{eqnarray}
\langle O \rangle^\prime &=&{\int {\cal D}\Phi O (\Phi_\vix )F^\prime (\Phi_\vix )\over{Z^\prime}}; 
\label{PQ_Z3_Potts} \\
F^\prime (\Phi_\vix )&=&e^{-S_{0,Z_3}-S_{{\rm R},Z_3}},~~~
Z^\prime =\int {\cal D}\Phi F^\prime (\Phi_\vix) .
\label{Z3_PQ_Potts_Z}
\end{eqnarray}
The phase factor is then given by  
\begin{eqnarray}
&&\langle {F\over{F^\prime}} \rangle^\prime =\langle e^{-iS_{{\rm I},Z_3}}\rangle^\prime ={\int {\cal D}\Phi 
[e^{-iS_{{\rm I},Z_3}}]F^\prime (\Phi_\vix )\over{Z^\prime}}
\nonumber\\
&&={\int {\cal D}\Phi 
[\cos{S_{{\rm I},Z_3}}]F^\prime (\Phi_\vix )\over{Z^\prime}}
=\langle \cos{S_{{\rm I},Z_3}}\rangle^\prime
={Z\over{Z^\prime}},   
\label{Z3_PQ_PF_Potts}
\end{eqnarray}
and the true average $\langle O \rangle$ is by 
\begin{eqnarray}
\langle O \rangle &=&{\int {\cal D}\Phi 
[O]F^\prime (\Phi_\vix )\over{Z}}={\langle 
O(\Phi )e^{-iS_{{\rm I},Z_3}}
\rangle^\prime \over{\langle {F\over{F^\prime}} \rangle^\prime }} 
\label{Z3_PQ_RW_Potts_2}
\end{eqnarray}

\begin{figure}[htbp]
\begin{center}
\includegraphics[width=0.5\textwidth]{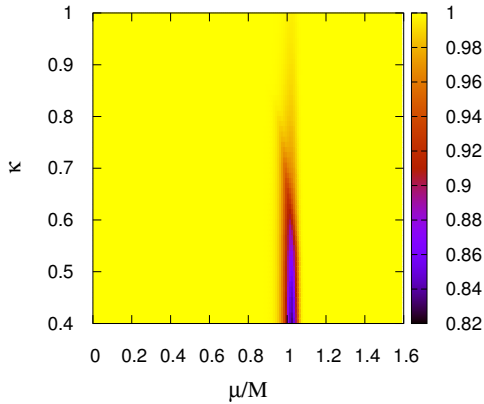}
\end{center}
\caption{The expectation value $\langle{\cos S_{{\mathrm I},Z_3}}\rangle^\prime$ calculated with the phase 
quenched approximation in $Z_3$-symmetric 3-d 13-state Potts model as a function of $\kappa$ and $\mu/M$. }
\label{Fig_Z3_13s_factor}
\end{figure}

Figure~\ref{Fig_Z3_13s_factor} shows the phase factor (\ref{Z3_PQ_PF_Potts}) in $\kappa$-$\mu$ plane.   
As an important result, the sign problem is serious only in the narrow region of $\mu/M \sim 1$ and $\kappa <0.9$. 

\begin{figure}[htbp]
\begin{center}
\includegraphics[width=0.5\textwidth]{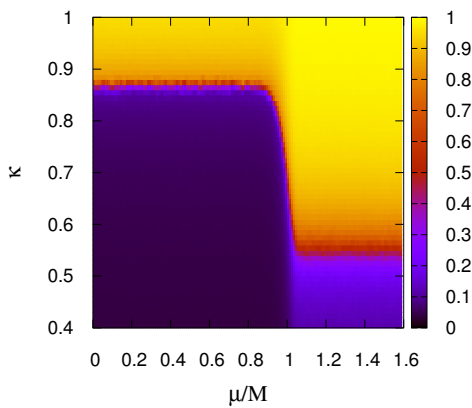}
\end{center}
\caption{The expectation value $\langle{|{\bar \Phi}|}\rangle$ calculated with the reweighting method in $Z_3$-symmetric 3-d 13-state Potts model as a function of $\kappa$ and $\mu/M$. 
}
\label{Fig_Z3_13s_diag}
\end{figure}

Figure~\ref{Fig_Z3_13s_diag} shows the expectation value of$\langle |{\bar \Phi}| \rangle$ in the $\kappa$-$\mu$ plane.  
The result was obtained by using the quenched approximated probability function and the reweighting method.   
The deconfinement transition line $\kappa=\kappa_{c}(\mu/M)$ is located 
 at $\kappa =0.87$ for $\mu/M < 1$, but goes down to $\kappa =0.55$ 
at $\mu/M \approx 1$, and keeps $\kappa =0.55$ for $\mu/M > 1$. 
Thus, $\mu$-dependence of $\kappa=\kappa_{c}(\mu/M)$ becomes strong as 
the number of states on each site increases. 
Eventually, the transition line of model (D) is rather similar to that of 
the original Potts model. 
Nevertheless, as for seriousness of the sign problem 
there is a big difference between the two models. 
Comparing Fig.~\ref{Fig_Z3_13s_factor} with Fig.~\ref{Fig_ordinary_factor}, 
we can conclude that the sign problem 
is almost cured by $Z_3$ symmetry, even if the value region of $\Phi_{\vix}$ is extended.

It is interesting that,  in Fig.~\ref{Fig_Z3_13s_factor}, the narrow dark region (where the phase factor is small) lies on the line $\mu/M \approx 1$ where the onset transition takes place. 
In fact, as shown in Fig.~\ref{Fig_n_density_kappa}, $n$ rapidly increases at $\mu/M \approx 1$. 
Of course, this dark region may be only a artifact due to the sign problem, but may imply that a physical sharp transition occurs 
on the line and the partition function $Z(\mu )$ becomes small around the line. 
Hence, Lee-Yang zeros analyses~\cite{Lee_Yang} may be valid in this phenomena. 
On the contrary, 
the deconfinement transition appears on the line of $\kappa \sim 0.87$ and 
$\mu/M < 1$ in Fig. \ref{Fig_Z3_13s_diag}, but 
any dark area is not found around the line in Fig.~\ref{Fig_Z3_13s_factor}. 
This may be caused by the fact that the lattice size is small in our 
simulations and the deconfinement transition does not induce 
any singular behavior in the partition function. 
This is an interesting question to be solved in future. 

\section{Summary}
\label{summary}

In summary, we have constructed four versions of $Z_3$-symmetric 3-d Potts 
models, each with different number of states, 
in order to investigate (1) the interplay between $Z_3$ symmetry 
and the sign problem and (2) the relation between 
the number of states and the deconfinement transition line 
in the $\mu$-$\kappa$ plane. 
Properties of the four $Z_3$-symmetric Potts models (A)-(D) 
are summarized in Table \ref{tb:models}, together with those of 
the original 3-d 3-state Potts model. 

As for subject (1), we have found 
from the comparison between the Potts model and model (D)
that the sign problem is almost cured by imposing $Z_3$ symmetry. 
The $Z_3$-symmetric Potts models are described by a power series of 
the Polyakov-loop operator $\Phi_{\vix}$ and its complex conjugate 
$\Phi_{\vix}^{*}$. $Z_3$ symmetry eliminates the linear term, 
so that the imaginary part of the model action 
starts with the terms of $(\Phi_{\vix})^3$ and $(\Phi_{\vix}^{*})^3$. 
This makes the sign problem less serious, because of $|\Phi_{\vix}| < 1$; 
note that only $Z_3$ 
elements satisfy $|\Phi_{\vix}| = 1$ but they do not contribute to 
the imaginary part. 
This mechanism may happen in $Z_3$-QCD. 
Therefore, there is a possibility that the sign problem is circumvented by 
the Taylor-expansion method of Ref.~\cite{Kouno_DFTBC} 
to derive QCD results from $Z_3$-QCD ones, even if $T\neq 0$. 
 
Subject (2) was clarified 
by changing the number of states from 3 of model (A) to 
13 of model (D). Comparing the results of model (A)-(D), 
we have found that $\mu$-dependence of the deconfinement transition line 
becomes stronger with respect to increasing the number of states. 

The lattice size we used is not so enough, 
so that there is possibility that the sign problem becomes more serious 
as the lattice volume becomes larger. 
We also postponed the determination of the order of 
the deconfinement transition for the same reason.
The study based on larger lattice is needed as a future work.  

There is a possibility that, as the original Potts model~\cite{Condella}, the $Z_3$ symmetric Potts model can be also transformed into the flux model that has no sign problem. 
In Ref.~\cite{Z3C}, the effective model based on the Polyakov-loop, the values of which are not restricted to $Z_3$ element, was transformed into the flux model. 
However, except for pure gauge term, only the linear terms of $\Phi_\vix$ are considered in that formalism.  
The extension of the formalism to the $Z_3$-symmetric Potts model is nontrivial, because the extension to the case with higher $\Phi_\vix$ terms is nontrivial. 
Therefore, we postpone the extension as a future problem.  

In the Potts models, chiral symmetry and its dynamical breaking cannot be discussed, since these models consider the case of 
large quark mass and thereby chiral symmetry is largely broken 
from the beginning. 
In QCD with light quark masses,  when the phase quenched approximation is used, 
chiral dynamics induces the problem of early onset of quark number 
density~\cite{Barbour} (or the baryon Silver Blaze problem~\cite{Cohen}). 
The problem may happen also in $Z_3$-QCD with light quark masses. 
Analyses beyond the phase quenched approximation may be important as a future problem.  
 
\noindent
\begin{acknowledgments}
The authors are thankful especially to  Hiroshi Suzuki and Hiroshi Yoneyama for crucial discussions on the realization of the QCD partition function.  
The authors also thank Atsushi Nakamura, Etsuko Itou, Masahiro Ishii, Junpei Sugano, Akihisa Miyahara, Shuichi Togawa and Yuhei Torigoe for fruitful discussions.  
H. K. also thanks Motoi Tachibana, Tatsuhiro Misumi, Yuya Tanizaki and Kouji Kashiwa for useful discussions. 
H. K. and M. Y. are supported by Grant-in-Aid for Scientific Research (No.26400279 and No.26400278) from Japan Society for the Promotion of Science (JSPS). 
The numerical calculations were partially performed by using SX-ACE at CMC, Osaka University.
\end{acknowledgments}



\begin{thebibliography}{19}
\expandafter\ifx\csname natexlab\endcsname\relax\def\natexlab#1{#1}\fi
\expandafter\ifx\csname bibnamefont\endcsname\relax
  \def\bibnamefont#1{#1}\fi
\expandafter\ifx\csname bibfnamefont\endcsname\relax
  \def\bibfnamefont#1{#1}\fi
\expandafter\ifx\csname citenamefont\endcsname\relax
  \def\citenamefont#1{#1}\fi
\expandafter\ifx\csname url\endcsname\relax
  \def\url#1{\texttt{#1}}\fi
\expandafter\ifx\csname urlprefix\endcsname\relax\def\urlprefix{URL }\fi
\providecommand{\bibinfo}[2]{#2}
\providecommand{\eprint}[2][]{\url{#2}}
%
\bibitem[{\citenamefont{Fodor}(2002)}]{Fodor}
\bibinfo{author}{\bibfnamefont{Z.}~\bibnamefont{Fodor}}, 
\bibnamefont{and} 
\bibinfo{author}{\bibfnamefont{S.}~\bibnamefont{D.}~\bibnamefont{Katz}},  
\bibinfo{journal}{Phys. Lett.\ B} \textbf{\bibinfo{volume}{534}}, 
\bibinfo{pages}{87} (\bibinfo{year}{2002}). 
%
\bibitem[{\citenamefont{Forcrand and Philipsen}(2002)}]{FP}
\bibinfo{author}{\bibfnamefont{P.}~\bibnamefont{de}~\bibnamefont{Forcrand}} 
\bibnamefont{and}
\bibinfo{author}{\bibfnamefont{O.}~\bibnamefont{Philipsen}},  
\bibinfo{journal}{Nucl. Phys. } \textbf{\bibinfo{volume}{B642}},
\bibinfo{pages}{290} (\bibinfo{year}{2002}). 
%
\bibitem[{\citenamefont{Elia and Lombardo}(2003)}]{D'Elia}
\bibinfo{author}{\bibfnamefont{M.}~\bibnamefont{D'Elia}} \bibnamefont{and}
\bibinfo{author}{\bibfnamefont{M.}~\bibfnamefont{P.}~\bibnamefont{Lombardo}},  
\bibinfo{journal}{Phys. Rev.\  D} \textbf{\bibinfo{volume}{67}},
\bibinfo{pages}{014505} (\bibinfo{year}{2003}). 
%
\bibitem[{\citenamefont{D'Elia et al}(2009)}]{D'Elia3}
\bibinfo{author}{\bibfnamefont{M.}~\bibnamefont{D'Elia}} \bibnamefont{and}
\bibinfo{author}{\bibfnamefont{F.}~\bibnamefont{Sanfilippo}},
\bibinfo{journal}{Phys.\ Rev.\ D} \textbf{\bibinfo{volume}{80}},
\bibinfo{pages}{111501} (\bibinfo{year}{2009}).
%
\bibitem[{\citenamefont{FP2010}(2009)}]{FP2010}
\bibinfo{author}{\bibfnamefont{P.}~\bibnamefont{de}~\bibnamefont
{Forcrand}} 
\bibnamefont{and}
\bibinfo{author}{\bibfnamefont{O.}~\bibnamefont{Philipsen}},  
\bibinfo{journal}{Phys.\ Rev.\ Lett. } \textbf{\bibinfo{volume}{105}},
\bibinfo{pages}{152001} (\bibinfo{year}{2010}). 
%
\bibitem{Nagata}
\bibinfo{author}{\bibfnamefont{K.}~\bibnamefont{Nagata}}
\bibnamefont{and}
\bibinfo{author}{\bibfnamefont{A.}~\bibnamefont{Nakamura}},  
\bibinfo{journal}{Phys. Rev. D} \textbf{\bibinfo{volume}{83}},
\bibinfo{pages}{114507} (\bibinfo{year}{2011}). 
%
\bibitem{Takahashi}
\bibinfo{author}{\bibfnamefont{J.}~\bibnamefont{Takahashi}}, 
\bibinfo{author}{\bibfnamefont{K.}~\bibnamefont{Nagata}}, 
\bibinfo{author}{\bibfnamefont{T.}~\bibnamefont{Saito}},
\bibinfo{author}{\bibfnamefont{A.}~\bibnamefont{Nakamura}}, 
\bibinfo{author}{\bibfnamefont{T.}~\bibnamefont{Sasaki}}, 
\bibinfo{author}{\bibfnamefont{H.}~\bibnamefont{Kouno}}, 
\bibnamefont{and} 
\bibinfo{author}{\bibfnamefont{M.}~\bibnamefont{Yahiro}} 
\bibinfo{journal}{Phys. Rev. D} \textbf{\bibinfo{volume}{88}},
\bibinfo{pages}{114504} (\bibinfo{year}{2013}); 
\bibinfo{author}{\bibfnamefont{J.}~\bibnamefont{Takahashi}}, 
\bibinfo{author}{\bibfnamefont{H.}~\bibnamefont{Kouno}}, 
\bibnamefont{and} 
\bibinfo{author}{\bibfnamefont{M.}~\bibnamefont{Yahiro}} 
\bibinfo{journal}{Phys. Rev. D} \textbf{\bibinfo{volume}{91}},
\bibinfo{pages}{014501} (\bibinfo{year}{2015}).
%
\bibitem[{\citenamefont{Allton}(2004)}]{Allton}
\bibinfo{author}{\bibfnamefont{C.}~\bibfnamefont{R.}~\bibnamefont{Allton}},
\bibinfo{author}{\bibfnamefont{S.}~\bibnamefont{Ejiri}},
\bibinfo{author}{\bibfnamefont{S.}~\bibfnamefont{J.}~\bibnamefont{Hands}},
\bibinfo{author}{\bibfnamefont{O.}~\bibnamefont{Kaczmarek}},
\bibinfo{author}{\bibfnamefont{F.}~\bibnamefont{Karsch}},
\bibinfo{author}{\bibfnamefont{E.}~\bibnamefont{Laermann}},
\bibinfo{author}{\bibfnamefont{Ch.}~\bibnamefont{Schmidt}},
\bibnamefont{and} 
\bibinfo{author}{\bibfnamefont{L.}~\bibnamefont{Scorzato}},
  \bibinfo{journal}{Phys. Rev. D} \textbf{\bibinfo{volume}{66}},
  \bibinfo{pages}{074507} (\bibinfo{year}{2002}). 
%
\bibitem[{\citenamefont{Ejiri et al.}(2004)}]{Ejiri_density}
\bibinfo{author}{\bibfnamefont{S.}~\bibnamefont{Ejiri}},
\bibinfo{author}{\bibfnamefont{Y.}~\bibnamefont{Maezawa}},
\bibinfo{author}{\bibfnamefont{N.}~\bibnamefont{Ukita}},
\bibinfo{author}{\bibfnamefont{S.}~\bibnamefont{Aoki}},
\bibinfo{author}{\bibfnamefont{T.}~\bibnamefont{Hatsuda}},
\bibinfo{author}{\bibfnamefont{N.}~\bibnamefont{Ishii}},
\bibinfo{author}{\bibfnamefont{K.}~\bibnamefont{Kanaya}},
\bibnamefont{and}
\bibinfo{author}{\bibfnamefont{T.}~\bibnamefont{Umeda}},
\bibinfo{journal}{Phys. Rev. D} \textbf{\bibinfo{volume}{82}},
\bibinfo{pages}{014508} (\bibinfo{year}{2010}). 
%
\bibitem{Aarts_CLE_1}
G. Aarts, 
\bibinfo{journal}{Phys.\ Rev.\ Lett. } \textbf{\bibinfo{volume}{102}},
\bibinfo{pages}{131601} (\bibinfo{year}{2009}). 
%
\bibitem{Aarts_CLE_2} 
G. Aarts, L. Bongiovanni, E. Seiler, D. Sexty, and I.-O. Stamatescu, 
\bibinfo{journal}{Eur.\ Phys.\ J.\ A } \textbf{\bibinfo{volume}{49}},
\bibinfo{pages}{89} (\bibinfo{year}{2013}). 
%
\bibitem{Sexty} 
D. Sexty,  
\bibinfo{journal}{Phys. Lett.\ B} \textbf{\bibinfo{volume}{729}},
\bibinfo{pages}{108} (\bibinfo{year}{2014}). 
%
\bibitem[{\citenamefont{Greensite}(2014)}]{Greensite}
\bibinfo{author}{\bibfnamefont{J.}~\bibnamefont{Greensite}}, 
arXiv:1406.4558 [hep-lat] (2014).  
%
\bibitem{Aarts_CLE_3} 
G. Aarts, F. Attanasio, B. J\"{a}ger, E. Seiler, D. Sexty, 
and I.-O. Stamatescu, arXiv:1411.2632 [hep-lat](2014).
%
\bibitem{Aurora_thimbles} 
M. Cristoforetti, F. Di Renzo, and L. Scorzato, 
\bibinfo{journal}{Phys.\ Rev.\ D } \textbf{\bibinfo{volume}{86}},
\bibinfo{pages}{074506} (\bibinfo{year}{2012}). 
%
\bibitem{Fujii_thimbles} 
H. Fujii, D. Honda, M. Kato, Y. Kikukawa, S. Komatsu and T. Sano, 
\bibinfo{journal}{JHEP} \textbf{\bibinfo{volume}{1310}},
\bibinfo{pages}{147} (\bibinfo{year}{2013}). 
%
\bibitem{Tanizaki}
Y. Tanizaki, H. Nishimura, K. Kashiwa,  
Phys. Rev. D 91, 101701 (2015). 
%
\bibitem[{\citenamefont{Condella and DeTar}(2000)}]{Condella}
\bibinfo{author}{\bibfnamefont{J.}~\bibnamefont{Condella}}, 
\bibnamefont{and}
\bibinfo{author}{\bibfnamefont{C.}~\bibnamefont{DeTar}},  
  \bibinfo{journal}{Phys.\ Rev.\  D} \textbf{\bibinfo{volume}{61}},
  \bibinfo{pages}{074023} (\bibinfo{year}{2000}).
%
\bibitem{Z3C}
Y.D. Mercado, H.G. Evertz and C.Gattringer, 
Phys. Rev. Lett. 106, (2011), 222001
;
C. Gattringer,
Nucl. Phys. B 850, (2011), 242 
;
Y.D. Mercado and C. Gattringer, 
Nucl. Phys. B 862, (2012),  737; 
Y.D. Mercado, H.G. Evertz and C. Gattringer,
Comput. Phys. Commun., 183, (2012), 1920. 
%
\bibitem{Langfeld_center}
K. Langfeld and A. Wipf, 
Annals of Phys. 327 (2012), 994. 
%
\bibitem{Z3A}
J. Bloch, F. Bruckmann and T. Wettig, 
JHEP 10 (2013) 140; 
J. Bloch, F. Bruckmann and T. Wettig, 
PoS(LATTICE 2013) 194, arXiv:1310.6645;  
J. Bloch and F. Bruckmann, 
arXiv:1508.03522.
%
\bibitem[{\citenamefont{Polyakov}(1978)}]{Polyakov}
\bibinfo{author}{\bibfnamefont{A.}~\bibfnamefont{M.}~\bibnamefont{Polyakov}}, 
\bibinfo{journal}{Phys. Lett.} \textbf{\bibinfo{volume}{72B}},
\bibinfo{pages}{477} (\bibinfo{year}{1978}).
%
\bibitem[{\citenamefont{Kouno et al}(2012)}]{Kouno_TBC}
\bibinfo{author}{\bibfnamefont{H.}~\bibnamefont{Kouno}}, 
\bibinfo{author}{\bibfnamefont{Y.}~\bibnamefont{Sakai}}, 
\bibinfo{author}{\bibfnamefont{T.}~\bibnamefont{Makiyama}},
\bibinfo{author}{\bibfnamefont{K.}~\bibnamefont{Tokunaga}},
\bibinfo{author}{\bibfnamefont{T.}~\bibnamefont{Sasaki}},   
\bibnamefont{and}
\bibinfo{author}{\bibfnamefont{M.}~\bibnamefont{Yahiro}},  
\bibinfo{journal}{J. Phys. G: Nucl. Part. Phys.} \textbf{\bibinfo{volume}{39}},
\bibinfo{pages}{085010} (\bibinfo{year}{2012}). 
%
\bibitem[{\citenamefont{Kouno et al}(2012)}]{Sakai_TBC}
\bibinfo{author}{\bibfnamefont{Y.}~\bibnamefont{Sakai}}, 
\bibinfo{author}{\bibfnamefont{H.}~\bibnamefont{Kouno}}, 
\bibinfo{author}{\bibfnamefont{T.}~\bibnamefont{Sasaki}},   
\bibnamefont{and}
\bibinfo{author}{\bibfnamefont{M.}~\bibnamefont{Yahiro}},  
\bibinfo{journal}{Phys. Lett.\ B} \textbf{\bibinfo{volume}{718}},
\bibinfo{pages}{130} (\bibinfo{year}{2012}). 
%
\bibitem[{\citenamefont{Kouno et al}(2013)}]{Kouno_adjoint}
\bibinfo{author}{\bibfnamefont{H.}~\bibnamefont{Kouno}}, 
\bibinfo{author}{\bibfnamefont{T.}~\bibnamefont{Misumi}},
\bibinfo{author}{\bibfnamefont{K.}~\bibnamefont{Kashiwa}},
\bibinfo{author}{\bibfnamefont{T.}~\bibnamefont{Makiyama}},
\bibinfo{author}{\bibfnamefont{T.}~\bibnamefont{Sasaki}},   
\bibnamefont{and}
\bibinfo{author}{\bibfnamefont{M.}~\bibnamefont{Yahiro}}, 
\bibinfo{journal}{Phys. Rev.\ D} \textbf{\bibinfo{volume}{88}},
\bibinfo{pages}{016002} (\bibinfo{year}{2013}). 
%
\bibitem[{\citenamefont{Kouno et al}(2013)}]{Kouno_TBC_2}
\bibinfo{author}{\bibfnamefont{H.}~\bibnamefont{Kouno}}, 
\bibinfo{author}{\bibfnamefont{T.}~\bibnamefont{Makiyama}},
\bibinfo{author}{\bibfnamefont{T.}~\bibnamefont{Sasaki}}, 
\bibinfo{author}{\bibfnamefont{Y.}~\bibnamefont{Sakai}},   
\bibnamefont{and}
\bibinfo{author}{\bibfnamefont{M.}~\bibnamefont{Yahiro}}, 
\bibinfo{journal}{J. Phys. G: Nucl. Part. Phys.} \textbf{\bibinfo{volume}{40}},
\bibinfo{pages}{095003} (\bibinfo{year}{2013}). 
%
\bibitem[{\citenamefont{Kouno et al}(2016)}]{Kouno_DFTBC}
\bibinfo{author}{\bibfnamefont{H.}~\bibnamefont{Kouno}}, 
\bibinfo{author}{\bibfnamefont{K.}~\bibnamefont{Kashiwa}},
\bibinfo{author}{\bibfnamefont{J.}~\bibnamefont{Takahashi}},
\bibinfo{author}{\bibfnamefont{T.}~\bibnamefont{Misumi}},   
\bibnamefont{and}
\bibinfo{author}{\bibfnamefont{M.}~\bibnamefont{Yahiro}}, 
\bibinfo{journal}{Phys. Rev.\ D} \textbf{\bibinfo{volume}{93}},
\bibinfo{pages}{056009} (\bibinfo{year}{2016}). 
%
\bibitem[{\citenamefont{Meisinger et al.}(1996)}]{Meisinger}
\bibinfo{author}{\bibfnamefont{P.}~\bibnamefont{N.}}~\bibnamefont{Meisinger},
\bibnamefont{and}
\bibinfo{author}{\bibfnamefont{M.}~\bibnamefont{C.}}~\bibnamefont{Ogilvie},  
  \bibinfo{journal}{Phys. Lett.\ B} \textbf{\bibinfo{volume}{379}},
  \bibinfo{pages}{163} (\bibinfo{year}{1996}). 
%
\bibitem[{\citenamefont{Dumitru}(2002)}]{Dumitru}
\bibinfo{author}{\bibfnamefont{A.}~\bibnamefont{Dumitru}},
\bibnamefont{and}
\bibinfo{author}{\bibfnamefont{R.}~\bibfnamefont{D.}~\bibnamefont{Pisarski}},  
\bibinfo{journal}{Phys.\ Rev.\  D} \textbf{\bibinfo{volume}{66}},
\bibinfo{pages}{096003} (\bibinfo{year}{2002}); 
\bibinfo{author}{\bibfnamefont{A.}~\bibnamefont{Dumitru}},
\bibinfo{author}{\bibfnamefont{Y.}~\bibnamefont{Hatta}},
\bibinfo{author}{\bibfnamefont{J.}~\bibnamefont{Lenaghan}},
\bibinfo{author}{\bibfnamefont{K.}~\bibnamefont{Orginos}},
\bibnamefont{and}
\bibinfo{author}{\bibfnamefont{R.}~\bibfnamefont{D.}~\bibnamefont{Pisarski}},  
\bibinfo{journal}{Phys.\ Rev.\  D} \textbf{\bibinfo{volume}{70}},
\bibinfo{pages}{034511} (\bibinfo{year}{2004}); 
\bibinfo{author}{\bibfnamefont{A.}~\bibnamefont{Dumitru}},
\bibinfo{author}{\bibfnamefont{R.}~\bibfnamefont{D.}~\bibnamefont{Pisarski}},  
\bibnamefont{and}
\bibinfo{author}{\bibfnamefont{D.}~\bibnamefont{Zschiesche}},  
\bibinfo{journal}{Phys.\ Rev.\  D} \textbf{\bibinfo{volume}{72}},
\bibinfo{pages}{065008} (\bibinfo{year}{2005}).
%
\bibitem[{\citenamefont{Fukushima}(2004)}]{Fukushima}
\bibinfo{author}{\bibfnamefont{K.}~\bibnamefont{Fukushima}}, 
  \bibinfo{journal}{Phys. Lett.\ B} \textbf{\bibinfo{volume}{591}},
  \bibinfo{pages}{277} (\bibinfo{year}{2004}). 
%
\bibitem[{\citenamefont{Ratti et al.}(2006)}]{Ratti}
\bibinfo{author}{\bibfnamefont{C.}~\bibnamefont{Ratti}},
\bibinfo{author}{\bibfnamefont{M.}~\bibfnamefont{A.}~\bibnamefont{Thaler}},
\bibnamefont{and}
\bibinfo{author}{\bibfnamefont{W.}~\bibnamefont{Weise}},  
  \bibinfo{journal}{Phys. Rev.\ D} \textbf{\bibinfo{volume}{73}},
  \bibinfo{pages}{014019} (\bibinfo{year}{2006}); 
\bibinfo{author}{\bibfnamefont{C.}~\bibnamefont{Ratti}},
\bibinfo{author}{\bibfnamefont{S.}~\bibnamefont{R\"{o}{\ss}ner}},
\bibinfo{author}{\bibfnamefont{M.}~\bibfnamefont{A.}~\bibnamefont{Thaler}},
\bibnamefont{and}
\bibinfo{author}{\bibfnamefont{W.}~\bibnamefont{Weise}},  
  \bibinfo{journal}{Eur. Phys. J.\ C} \textbf{\bibinfo{volume}{49}},
  \bibinfo{pages}{213} (\bibinfo{year}{2007}). 
%
\bibitem[{\citenamefont{Megias al.}(2006)}]{Megias}
\bibinfo{author}{\bibfnamefont{E.}~\bibnamefont{Megias}}, 
\bibinfo{author}{\bibfnamefont{E.~R.}~\bibnamefont{ Arriola}}, 
\bibnamefont{and}
\bibinfo{author}{\bibfnamefont{L.~L.}~\bibnamefont{ Salcedo}},
\bibinfo{journal}{Phys.\ Rev.\ D} \textbf{\bibinfo{volume}{74}},
\bibinfo{pages}{065005} (\bibinfo{year}{2006}).
%
\bibitem{Iritani}
T. Iritani, E. Itou, T. Misumi,  
arXiv:1508.07132, to apper in JHEP; 
T. Misumi, T. Iritani, E. Itou,  
presented at the 33rd International Symposium on Lattice Field Theory, Lattice2015, 14-18 July 2015, Kobe International Conference Center, Kobe, JAPAN, 
 arXiv:1510.07227. 
%
\bibitem{Greensite:2014cxa} 
  J.~Greensite,
  Phys.\ Rev.\ D {\bf 90}, no. 11, 114507 (2014)
  doi:10.1103/PhysRevD.90.114507
  [arXiv:1406.4558 [hep-lat]]. 
%
\bibitem{Aarts:2014kja} 
  G.~Aarts, F.~Attanasio, B.~Jager, E.~Seiler, D.~Sexty and I.~O.~Stamatescu,
  PoS LATTICE {\bf 2014}, 200 (2014)
  [arXiv:1411.2632 [hep-lat]].
%
\bibitem[{\citenamefont{DeGrand and DeTar}(1983)}]{DeGrand}
\bibinfo{author}{\bibfnamefont{T.A.}~\bibnamefont{DeGrand}} \bibnamefont{and}
\bibinfo{author}{\bibfnamefont{C.E.}~\bibnamefont{DeTar}},  
\bibinfo{journal}{Nucl. Phys. } \textbf{\bibinfo{volume}{B225}},
\bibinfo{pages}{590} (\bibinfo{year}{1983}). 
%
\bibitem[{\citenamefont{Karsch and Stickan}(2000)}]{Karsch}
\bibinfo{author}{\bibfnamefont{F.}~\bibnamefont{Karsch}} 
\bibnamefont{and}
\bibinfo{author}{\bibfnamefont{S.}~\bibnamefont{Stickan}},  
\bibinfo{journal}{Phys. Lett. \ B} \textbf{\bibinfo{volume}{488}},
\bibinfo{pages}{319} (\bibinfo{year}{2000}). 
%
\bibitem{Alford}
\bibinfo{author}{\bibfnamefont{M.}~\bibnamefont{Alford}},
\bibinfo{author}{\bibfnamefont{S.}~\bibnamefont{Chandrasekharan}}
\bibinfo{author}{\bibfnamefont{J.}~\bibnamefont{Cox}}
\bibnamefont{and}
\bibinfo{author}{\bibfnamefont{U.-J.}~\bibnamefont{Wiese}}, 
\bibinfo{journal}{Nucl. Phys.} \textbf{\bibinfo{volume}{B602}},
\bibinfo{pages}{61} (\bibinfo{year}{2001}). 
%
\bibitem[{\citenamefont{Fujii}(2003)}]{Fujii}
\bibinfo{author}{\bibfnamefont{H.}~\bibnamefont{Fujii}}, 
\bibinfo{journal}{Phys.\ Rev.\ D} \textbf{\bibinfo{volume}{67}},
\bibinfo{pages}{094018} (\bibinfo{year}{2003}); 
\bibinfo{author}{\bibfnamefont{H.}~\bibnamefont{Fujii}},  
\bibnamefont{and}
\bibinfo{author}{\bibfnamefont{M.}~\bibnamefont{ Ohtani}},
\bibinfo{journal}{Phys.\ Rev.\ D} \textbf{\bibinfo{volume}{70}},
\bibinfo{pages}{014016} (\bibinfo{year}{2004}).
%
\bibitem{Bjorken}
J. D. Bjorken and S. D. Drell, "Relativistic Quantum Fields", McGraw-Hill, 1965.
%
\bibitem[{\citenamefont{Roberge and Weiss}(1986)}]{RW}
\bibinfo{author}{\bibfnamefont{A.}~\bibnamefont{Roberge}} \bibnamefont{and}
\bibinfo{author}{\bibfnamefont{N.}~\bibnamefont{Weiss}},  
\bibinfo{journal}{Nucl. Phys. } \textbf{\bibinfo{volume}{B275}},
\bibinfo{pages}{734} (\bibinfo{year}{1986}). 
%
\bibitem[{\citenamefont{McLerran and Pisarski}(2007)}]{MacLerran}
\bibinfo{author}{\bibfnamefont{L.}~\bibnamefont{McLerran}} \bibnamefont{and}
\bibinfo{author}{\bibfnamefont{R.D.}~\bibnamefont{Pisarski}},  
\bibinfo{journal}{Nucl. Phys. } \textbf{\bibinfo{volume}{A796}},
\bibinfo{pages}{83} (\bibinfo{year}{2007}). 
%
\bibitem[{\citenamefont{Hidaka, McLerran and Pisarski}(2008)}]{Hidaka}
\bibinfo{author}{\bibfnamefont{Y.}~\bibnamefont{Hidaka}},
\bibinfo{author}{\bibfnamefont{L.}~\bibnamefont{McLerran}} \bibnamefont{and}
\bibinfo{author}{\bibfnamefont{R.D.}~\bibnamefont{Pisarski}},  
\bibinfo{journal}{Nucl. Phys. } \textbf{\bibinfo{volume}{A808}},
\bibinfo{pages}{117} (\bibinfo{year}{2008}). 
%
\bibitem[{\citenamefont{Lee-Yang}(1952)}]{Lee_Yang}
\bibinfo{author}{\bibfnamefont{C.N.}~\bibnamefont{Yang}}, 
\bibnamefont{and}
\bibinfo{author}{\bibfnamefont{T.D.}~\bibnamefont{Lee}},  
\bibinfo{journal}{Phys.\ Rev.} \textbf{\bibinfo{volume}{87}},
\bibinfo{pages}{404} (\bibinfo{year}{1952}); 
\bibinfo{author}{\bibfnamefont{T.D.}~\bibnamefont{Lee}},  
\bibnamefont{and}
\bibinfo{author}{\bibfnamefont{C.N.}~\bibnamefont{Yang}},
\bibinfo{journal}{Phys.\ Rev.} \textbf{\bibinfo{volume}{87}},
\bibinfo{pages}{410} (\bibinfo{year}{1952}).
%
\bibitem[{\citenamefont{Barbour et al.}(1999)}]{Barbour}
\bibinfo{author}{\bibfnamefont{I.M.}~\bibnamefont{Barbour}},
\bibnamefont{and}
\bibinfo{author}{\bibfnamefont{A.J.}~\bibnamefont{Bell}},  
\bibinfo{journal}{Nucl. Phys. } \textbf{\bibinfo{volume}{B372}},
\bibinfo{pages}{385} (\bibinfo{year}{1992}); 
\bibinfo{author}{\bibfnamefont{I.M.}~\bibnamefont{Barbour}},
\bibinfo{author}{\bibfnamefont{S.E.}~\bibnamefont{Morrison}},  
\bibinfo{author}{\bibfnamefont{E.G.}~\bibnamefont{Klepfish}}, 
\bibinfo{author}{\bibfnamefont{J.B. }~\bibnamefont{Kogut}},
\bibnamefont{and}
\bibinfo{author}{\bibfnamefont{M.-P.}~\bibnamefont{Lombardo}},  
\bibinfo{journal}{Phys. Rev. \ D} \textbf{\bibinfo{volume}{56}},
\bibinfo{pages}{7063} (\bibinfo{year}{1997}); 
\bibinfo{author}{\bibfnamefont{I.M.}~\bibnamefont{Barbour}},
\bibinfo{author}{\bibfnamefont{S.E.}~\bibnamefont{Morrison}},  
\bibinfo{author}{\bibfnamefont{E.G.}~\bibnamefont{Klepfish}}, 
\bibinfo{author}{\bibfnamefont{J.B. }~\bibnamefont{Kogut}},
\bibnamefont{and}
\bibinfo{author}{\bibfnamefont{M.-P.}~\bibnamefont{Lombardo}},  
\bibinfo{journal}{Nucl. Phys. Prcc. Suppl. } \textbf{\bibinfo{volume}{60A}},
\bibinfo{pages}{220} (\bibinfo{year}{1998}); 
\bibinfo{author}{\bibfnamefont{I.}~\bibnamefont{Barbour}},
\bibinfo{author}{\bibfnamefont{S.}~\bibnamefont{Hands}}, 
\bibinfo{author}{\bibfnamefont{J.B. }~\bibnamefont{Kogut}},
\bibinfo{author}{\bibfnamefont{M.-P.}~\bibnamefont{Lombardo}},  
\bibnamefont{and}
\bibinfo{author}{\bibfnamefont{S.}~\bibnamefont{Morrison}},  
\bibinfo{journal}{Nucl. Phys. } \textbf{\bibinfo{volume}{B557}},
\bibinfo{pages}{327} (\bibinfo{year}{1999}). 
%
\bibitem[{\citenamefont{Cohen}(2003)}]{Cohen}
\bibinfo{author}{\bibfnamefont{T.D.}~\bibnamefont{Cohen}},  
\bibinfo{journal}{Phys. Rev. Lett.} \textbf{\bibinfo{volume}{91}},
\bibinfo{pages}{222001} (\bibinfo{year}{2003}). 
%



\end{thebibliography}
\end{document}